\documentclass[sigconf,nonacm]{acmart}

\usepackage{comment}
\usepackage{amsmath}
\usepackage{listings}

\usepackage{transparent}
\usepackage{url}
\usepackage{xcolor}
\usepackage{subcaption}
\usepackage{cuted}

\usepackage{algorithm}
\usepackage[noend]{algpseudocode}

\usepackage[apply]{xnotes}

\usepackage{bbding}

\usepackage{hyperref}
\hypersetup{linkbordercolor=green}

\newtheorem{theorem}{Theorem}[section]

\newtheorem{definition}[theorem]{Definition}
\newtheorem{lemma}[theorem]{Lemma}

\definecolor{medgreen}{RGB}{0,150,0}
\definecolor{darkgreen}{RGB}{0,120,0}

\AddXNotesUser{pk}{PK}{orange}
\AddXNotesUser{at}{AT}{medgreen}

\newenvironment{executionexample}%
{%

    \newcommand{\textevil}{\includeemoji[1.2]{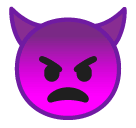}}
    
    \newcommand{\textlock}{\includeemoji[1.2]{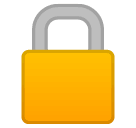}}}
{%

    \let\textevil\undefined
    
    \let\textlock\undefined}

\algblock[Block]{BeginBlock}{EndBlock}
\algtext*{BeginBlock}
\algtext*{EndBlock}

\newcommand\SharedObjects{\Statex \textbf{Shared objects:} \BeginBlock}
\newcommand\EndSharedObjects{\EndBlock}

\newcommand\GlobalsBlockName{Global variables}

\newcommand\Globals{\Statex \textbf{\GlobalsBlockName:} \BeginBlock}
\newcommand\EndGlobals{\EndBlock}

\newcommand\AuxiliaryFunctionsInline{\Statex \textbf{Auxiliary functions:\ }}
\newcommand\AuxiliaryFunctions{\Statex \textbf{Auxiliary functions:} \BeginBlock}
\newcommand\EndAuxiliaryFunctions{\EndBlock}

\newcommand\ParametersInline{\Statex \textbf{Parameters:\ }}
\newcommand\Parameters{\Statex \textbf{Parameters:} \BeginBlock}
\newcommand\EndParameters{\EndBlock}

\newcommand\Interface{\Statex \textbf{Interface:} \BeginBlock}
\newcommand\EndInterface{\EndBlock}

\newcommand\PropertiesInline{\Statex \textbf{Properties:\ }}

\algblockdefx[UponReceive]{UponReceive}{EndHandler}[2]{\textbf{upon receive} $\langle$#1$\rangle$ \textbf{from} #2}{}
\algtext*{EndHandler}  

\algblockdefx[UponRBDeliver]{UponRBDeliver}{EndHandler}[2]{\textbf{upon RB-deliver} $\langle$#1$\rangle$ \textbf{from} #2}{}
\algtext*{EndHandler}  

\algblockdefx[UponURBDeliverIn]{UponURBDeliverIn}{EndHandler}[3]{\textbf{upon URB-deliver} $\langle$#1$\rangle$ \textbf{in} #2 \textbf{from} #3}{}
\algtext*{EndHandler}  

\algblockdefx[Upon]{Upon}{EndHandler}[1]{\textbf{upon} #1}{}
\algtext*{EndHandler}  

\algblockdefx[ForEach]{ForEach}{EndFor}[1]{\textbf{for each} #1 \textbf{do}}{}
\algtext*{EndFor}  

\algblockdefx[Routine]{Routine}{EndRoutine}[1]{\textbf{routine} #1}{}
\algtext*{EndRoutine}  

\algblockdefx[Operation]{Operation}{EndOperation}[2]{\textbf{operation} #1(#2)}{}
\algtext*{EndOperation}  

\algblockdefx[Upcall]{Upcall}{EndUpcall}[2]{\textbf{upcall} #1(#2)}{}
\algtext*{EndUpcall}  

\algtext*{EndIf}
\algtext*{EndFunction}
\algtext*{EndProcedure}

\newcommand\Tupple[1]{\langle#1\rangle}
\newcommand\Message[1]{\ensuremath{\langle}#1\ensuremath{\rangle}}
\newcommand\MsgDesc[1]{\normalfont\textbf{#1}}

\newcommand\Send[2]{\textbf{send} $\langle$#1$\rangle$ \textbf{to} #2}
\newcommand\RBBroadcast[1]{\textbf{RB-Broadcast} $\langle$#1$\rangle$}
\newcommand\URBBroadcastIn[2]{\textbf{URB-Broadcast} $\langle$#1$\rangle$ \textbf{in} #2}
\newcommand\WaitFor{\textbf{wait for }}

\algnewcommand{\LeftComment}[1]{\(\triangleright\) #1}
\algnewcommand{\LineComment}[1]{\State \LeftComment{#1}}
\algnewcommand{\LineCommentx}[1]{\Statex \LeftComment{#1}}

\makeatletter
\newenvironment{pagealgorithm*}
    {
        \begin{strip}
        \refstepcounter{algorithm}
        \hrule height.8pt depth0pt \kern2pt
        \renewcommand{\caption}[2][\relax]{
            {\textbf{\fname@algorithm~\thealgorithm} ##2\par}%
            \ifx\relax##1\relax 
                \addcontentsline{loa}{algorithm}{\protect\numberline{\thealgorithm}##2}%
            \else 
                \addcontentsline{loa}{algorithm}{\protect\numberline{\thealgorithm}##1}%
            \fi
            \kern2pt\hrule\kern2pt
        }
    }{
        \kern2pt\hrule\relax
        \end{strip}
        \newpage 
    }
\makeatother

\newcommand\suchThat{\;\text{such that}\;}
\newcommand\textST{\;\text{s.t.}\;}
\newcommand\bydef{\triangleq}

\newcommand\attackName{``I still work here''}

\newcommand\pBLAVerifiability{BLA-Verifiability}
\newcommand\pBLAValidity{BLA-Validity}
\newcommand\pBLAComparability{BLA-Comparability}
\newcommand\pBLAInclusion{BLA-Inclusion}
\newcommand\pBLALiveness{BLA-Liveness}

\newcommand\pReconfigurationValidity{Reconfiguration Validity}
\newcommand\pReconfigurationLiveness{Reconfiguration Liveness}
\newcommand\pDynamicValidity{Dynamic Validity}
\newcommand\pDynamicLiveness{Dynamic Liveness}
\newcommand\pInstallationLiveness{Installation Liveness}

\newcommand\pCertificateVerifiability{Certificate Verifiability}

\newcommand\QuorumIntersection{quorum intersection}

\newcommand\pMRValidity{MR-Validity}
\newcommand\pMRAtomicity{MR-Atomicity}
\newcommand\pMRLiveness{MR-Liveness}

\newcommand\Replicas{\mathit{\Phi}}
\newcommand\Clients{\mathit{\Pi}}
\newcommand\Updates{\mathit{Updates}}
\newcommand\Lat{\mathcal{L}}
\newcommand\Hist{\mathcal{H}}
\newcommand\Conf{\mathcal{C}}
\newcommand\Values{\mathbb{V}}

\newcommand\Vinit{\mathit{Vinit}}
\newcommand\Cinit{\mathit{Cinit}}
\newcommand\Cnext{\mathit{Cnext}}
\newcommand\Ccurr{\mathit{Ccurr}}
\newcommand\Cinst{\mathit{Cinst}}
\newcommand\Chighest{\mathit{Chighest}}
\newcommand\Cmax{\mathit{Cmax}}
\newcommand\Sigmahistory{\mathit{\sigma_{history}}}

\newcommand\Cnice{\widetilde{\mathcal{C}}}

\newcommand\thigher{higher}

\newcommand\thighest{highest}

\newcommand\tlower{lower}

\newcommand\tlowest{lowest}

\newcommand\pivotal{pivotal}

\newcommand\tentative{tentative}
\newcommand\Tentative{Tentative}
\newcommand\tactive{active}
\newcommand\tsuperseded{superseded}

\newcommand\dynamic{dynamic}
\newcommand\Dynamic{Dynamic}
\newcommand\reconfigurable{reconfigurable}
\newcommand\Reconfigurable{Reconfigurable}

\newcommand\AccessControl{Access Control}
\newcommand\DynamicAccessControl{{\Dynamic} Access Control}
\newcommand\ReconfigurableAccessControl{{\Reconfigurable} Access Control}
\newcommand\VoteBasedDynamicAccessControl{Vote-Based {\Dynamic} Access Control}

\newcommand\LatticeAgreement{Lattice Agreement}
\newcommand\GeneralizedLatticeAgreement{Generalized Lattice Agreement}
\newcommand\ByzantineLatticeAgreement{Byzantine Lattice Agreement}
\newcommand\BLA{BLA}
\newcommand\DynamicByzantineLatticeAgreement{Dynamic Byzantine Lattice Agreement}
\newcommand\DBLA{DBLA}
\newcommand\ReconfigurableByzantineLatticeAgreement{Reconfigurable Byzantine Lattice Agreement}

\newcommand\MaxRegister{Max-Register}
\newcommand\DynamicMaxRegister{{\Dynamic} Max-Register}
\newcommand\DMR{DMR}

\newcommand\iTrue{\mathit{true}}
\newcommand\iFalse{\mathit{false}}
\newcommand\tlaStyleLand{$\State\qquad$\land\;}
\newcommand\tlaStyleLor{$\State\qquad$\lor\;}

\newcommand\ireplicas{\mathit{replicas}}

\newcommand\ihistory{\mathit{history}}
\newcommand\iprocessId{\mathit{processId}}
\newcommand\isignature{\mathit{sig}}
\newcommand\ivalues{\mathit{vs}}
\newcommand\istatus{\mathit{status}}
\newcommand\iinactive{\mathit{inactive}}
\newcommand\iproposing{\mathit{proposing}}
\newcommand\iconfirming{\mathit{confirming}}

\newcommand\iacks{\mathit{acks}}
\newcommand\iproposeAcks{\mathit{proposeAcks}}
\newcommand\iconfirmAcks{\mathit{confirmAcks}}
\newcommand\icurVals{\mathit{curVals}}
\newcommand\icurrentValues{\icurVals}

\newcommand\iConfRAC{\mathit{ConfRAC}}
\newcommand\iConfLA{\mathit{ConfLA}}
\newcommand\iHistLA{\mathit{HistLA}}

\newcommand\iDObj{\mathit{DObj}}

\newcommand\ist{\mathit{st}}
\newcommand\isk{\mathit{sk}}

\newcommand\iquorums{\mathit{quorums}}
\newcommand\ivs{\mathit{vs}}

\newcommand\iseqNumber{\mathit{seqNum}}
\newcommand\isn{\mathit{sn}}

\newcommand\isigValid{\mathit{sigValid}}
\newcommand\isuccess{\mathit{success}}
\newcommand\ireadOk{\mathit{readOk}}

\newcommand{\iactiveStateTransfer}{\mathit{inStateTransfer}}

\newcommand\vcurr{v_{curr}}
\newcommand\sigmacurr{\sigma_{curr}}

\newcommand{\FunctionName}[1]{\textsc{#1}}
\newcommand\fMaxElement{\FunctionName{HighestConf}}
\newcommand\fPropose{\FunctionName{Propose}}

\newcommand\fUpdateConfig{\FunctionName{UpdateConfig}}

\newcommand\fRefine{\FunctionName{Refine}}

\newcommand\fUpdateHistory{\FunctionName{UpdateHistory}}

\newcommand\fVerifyValues{\FunctionName{VerifyInputValues}}
\newcommand\fVerifyInputValue{\FunctionName{VerifyInputValue}}
\newcommand\fVIV{\FunctionName{VIV}}

\newcommand\fVerifyInputConfig{\FunctionName{VerifyInputConfig}}
\newcommand\fVIC{\FunctionName{VIC}}

\newcommand\fVerifyHistory{\FunctionName{VerifyHistory}}

\newcommand\fVH{\FunctionName{VH}}

\newcommand\fFSVerify{\FunctionName{FSVerify}}
\newcommand\fFSSign{\FunctionName{FSSign}}
\newcommand\fUpdateFSKeys{\FunctionName{UpdateFSKey}}

\newcommand\fContainsQuorum{\FunctionName{ContainsQuorum}}
\newcommand\fJoinAll{\FunctionName{JoinAll}}
\newcommand\iheight{\mathit{height}}
\newcommand\fVerifyOutputValue{\FunctionName{VerifyOutputValue}}

\newcommand\fRead{\FunctionName{Read}}
\newcommand\fWrite{\FunctionName{Write}}
\newcommand\fGet{\FunctionName{Get}}
\newcommand\fSet{\FunctionName{Set}}

\newcommand\fInstalledConfig{\FunctionName{InstalledConfig}}

\newcommand\fRequestCert{\FunctionName{RequestCert}}
\newcommand\fVerifyCert{\FunctionName{VerifyCert}}
\newcommand\fVoteYes{\FunctionName{VoteYes}}

\newcommand\mUpdateRead{\MsgDesc{UpdateRead}}
\newcommand\mUpdateReadResp{\MsgDesc{UpdateReadResp}}

\newcommand\mUpdateComplete{\MsgDesc{UpdateComplete}}

\newcommand\mPropose{\MsgDesc{Propose}}
\newcommand\mProposeResp{\MsgDesc{ProposeResp}}
\newcommand\mConfirm{\MsgDesc{Confirm}}
\newcommand\mConfirmResp{\MsgDesc{ConfirmResp}}
\newcommand\mNewHistory{\MsgDesc{NewHistory}}

\newcommand\mRequest{\MsgDesc{Request}}
\newcommand\mYes{\MsgDesc{Yes}}
\newcommand\mNo{\MsgDesc{No}}

\newcommand\mGet{\MsgDesc{Get}}
\newcommand\mGetResp{\MsgDesc{GetResp}}
\newcommand\mSet{\MsgDesc{Set}}
\newcommand\mSetResp{\MsgDesc{SetResp}}

\newlength{\BHeight}

\newcommand\algspace{\vspace{0.4\baselineskip}}

\newcounter{myLineNumber}
\def\BeginAlgorithmic{
\begin{algorithmic}[1]
\makeatletter
\setcounter{ALG@line}{\value{myLineNumber}}
\makeatother
}

\def\EndAlgorithmic{
\makeatletter
\setcounter{myLineNumber}{\value{ALG@line}}
\makeatother
\end{algorithmic}
}

\newcommand\algname{}
\newcommand\labelalg[1]{\renewcommand\algname{#1} \label{alg:\algname}}
\newcommand\labelalgpart[2]{\renewcommand\algname{#1} \label{alg:\algname-part-#2}}
\newcommand\labelline[1]{\label{lst:\algname:#1}}
\newcommand\lstref[1]{\ref{alg:#1}}
\newcommand\lstlineref[2]{line~\ref{lst:#1:#2}}
\newcommand\lstlinerangeref[3]{lines~\ref{lst:#1:#2}--\ref{lst:#1:#3}}

\newcommand{\myparagraph}[1]{\paragraph*{#1}}
\newcommand{\needsrev}[1]{#1}

\newcommand{\confirmA}[1]{#1}

\newcommand{\oldconfirm}[1]{#1}

\graphicspath{{compiled-drawings/}}
\newcommand\pdftex[2][]{

    \ifthenelse{\equal{#1}{}}{}{\def\svgwidth{#1}}
    \input{#2}
}

\begin{document}

\title{Asynchronous Reconfiguration with Byzantine Failures}

\author{Petr Kuznetsov}
\email{petr.kuznetsov@telecom-paris.fr}
\affiliation{%
  \institution{LTCI, T\'el\'ecom Paris, Institut Polytechnique de Paris}
  \country{France}
}

\author{Andrei Tonkikh}
\email{andrei.tonkikh@gmail.com}
\affiliation{%
  \institution{HSE University}
  \country{Russia}
}

\begin{abstract}
    Replicated services are inherently vulnerable to failures and security
    breaches.
    In a long-running system, it is, therefore, indispensable to maintain a
    \emph{reconfiguration} mechanism that would replace faulty
    replicas with correct ones.
    %
    %
    An important challenge is to enable reconfiguration without
    affecting the availability and consistency of the replicated data:
    the clients should be able to get correct service even when the set
    of service replicas is being updated.

    In this paper, we address the problem of reconfiguration in the
    presence of Byzantine failures: faulty replicas or clients may
    arbitrarily deviate from their expected behavior.
    We describe a generic technique for building  \emph{asynchronous} and \emph{Byzantine fault-tolerant}
    reconfigurable objects: clients can manipulate the object data and
    issue reconfiguration calls without reaching consensus on the
    current configuration.
    With the help of forward-secure digital signatures,
    our solution makes sure that superseded and possibly compromised configurations are harmless,
    that slow clients cannot be fooled into reading stale data,
    and that Byzantine clients cannot cause a denial of service
    by flooding the system with reconfiguration requests.
    %
    Our approach is modular and based on \emph{{\dynamic} \needsrev{Byzantine} lattice agreement} abstraction,
    and we discuss how to extend it to enable Byzantine fault-tolerant implementations of a large class of reconfigurable replicated services.
\end{abstract}


\maketitle


\section{Introduction}\label{sec:introduction}

\myparagraph{Replication and quorums.}
Replication is a natural way to ensure availability of shared data in the
presence of failures.
A collection of \emph{replicas}, each holding a version
of the data, ensure that the \emph{clients} get
a desired service, even when some replicas become unavailable or
hacked by a malicious adversary.
Consistency of the provided service requires the replicas to
\emph{synchronize}:
intuitively, every client should be able to
operate on the most ``up-to-date'' data, regardless of the set of
replicas it can reach.

It always makes sense to assume as little as possible about the
environment in which a system we design  is expected to run.
For example, \emph{asynchronous}  distributed systems do not rely on timing
assumptions, which makes them extremely robust with respect to communication
disruptions and computational delays.
It is, however, notoriously difficult and sometimes even impossible
to make such systems \emph{fault-tolerant}.
The folklore CAP theorem~\cite{Bre00,GL02} states that no replicated service can
combine consistency, availability, and partition-tolerance.
In particular, no consistent and available read-write storage can be
implemented in the presence of partitions: clients in one partition
are unable to keep track of the updates taking place in another one.

Therefore, fault-tolerant storage systems tend to assume that partitions are
excluded, e.g., by requiring a majority of replicas to be \emph{correct}~\cite{ABD}.
More generally, one can assume a \emph{quorum system}, e.g., a set of
subsets of replicas satisfying  the intersection and availability properties~\cite{quorums}.
Every (read or write) request from a client should be \emph{accepted}
by a quorum of replicas.
As every two quorums have at least one replica in common, intuitively,
no client can miss previously written data.

Of course, failures of replicas may jeopardize the underlying quorum
system.
In particular, we may find ourselves in a system in which no
quorum is available and, thus, no operation may be able to terminate.
Even worse, if the replicas are subject to Byzantine failures, we may
not be able to guarantee the very correctness of read values.

\myparagraph{Asynchronous reconfiguration.}
To anticipate such scenarios in a long run, we must maintain a
\emph{reconfiguration} mechanism that enables replacing compromised
replicas with correct ones and update the corresponding quorum assumptions.
A challenge here is to find an asynchronous implementation of
reconfiguration in a system where both clients and replicas are
subject to Byzantine failures that can be manifested by arbitrary and even malicious behavior.
In the world of selfishly driven blockchain users, a reconfiguration mechanism must be prepared for this.

Recently, a number of reconfigurable systems were
proposed for asynchronous \emph{crash}-fault
environments~\cite{dynastore,parsimonious,smartmerge,freestore,SKM17-reconf,rla}
that were first applied to (read-write) storage systems~\cite{dynastore,
  parsimonious,freestore}, and then extended to
max-registers~\cite{smartmerge,SKM17-reconf} and more general
\emph{lattice} data type~\cite{rla}.

These proposals tend to ensure that the clients
reach a form of ``loose'' agreement on the currently active
configurations, which can be naturally expressed via the \emph{lattice
  agreement} abstraction~\cite{lattice-hagit,gla}.
We allow clients to (temporarily) live in different worlds, as long as these
worlds are properly ordered.
For example, we may represent a configuration as a set of \emph{updates} (additions and removals of replicas)
and require that all installed configurations should be
related by containment.
A configuration becomes \emph{stale} as soon as a new configuration
representing a proper superset of updates is installed.

\myparagraph{Challenges of Byzantine fault-tolerant reconfiguration.}
In this paper, \needsrev{we focus on \emph{Byzantine fault-tolerant} reconfiguration.}
We have to address here several challenges, specific to dynamic systems with
Byzantine faults, \needsrev{which does not allow to simply employ the existing crash fault-tolerant solutions.}

First, when we build a system out of lower-level components, we need
to make sure that the outputs provided by these components are ``authentic''.
Whenever a (potentially Byzantine) process claims to have obtained a
\emph{value} $v$ (e.g., a new configuration estimate) from an underlying object (e.g., Lattice Agreement), it should
also provide a \emph{proof} $\sigma$ that can be independently verified by every
correct process.
The proof typically consists of digital signatures provided by a quorum of replicas of some configuration.
We abstract this requirement out by equipping the object with a function
$\fVerifyOutputValue$ that returns a boolean value, provided $v$ and $\sigma$.
When invoked by a correct process, the function returns
$\iTrue$ if and only if $v$ has indeed been produced by the
object.
When ``chaining'' the objects, i.e., adopting the output $v$ provided by an
object $A$ as an input for another object $B$, which is the typical scenario in
our system, a correct process invokes
$A.\fVerifyOutputValue(v,\sigma)$, where $\sigma$ is the proof
associated with $v$ by the implementation of $A$.
This way, only values actually produced by $A$ can be used as inputs to $B$.

Second, we face the \textbf{``I still work here'' attack}~\cite{tutorial2010}.
It is possible that a client that did not log into the system for a long time tries to access a stale
configuration in which some quorum is entirely compromised by the
Byzantine adversary.
The client can therefore be provided with an inconsistent view on the
shared data.
Thus, before accepting a new configuration, we need to make sure
that the stale ones are no longer capable of processing data requests from the clients.
We address this issue via the use of a \emph{forward-secure} signature
scheme~\cite{bellare1999forward}.
Intuitively, every replica is provided with a distinct private key
associated to each configuration.
Before a configuration is replaced with a newer one,
at least a quorum of its replicas are asked to destroy their private keys.
Therefore, even if the replicas are to become Byzantine in the future,
they will not be able to provide slow clients with inconsistent
values.
The stale configuration simply becomes non-responsive, as in crash-fault-tolerant reconfigurable systems.

Unfortunately, in an asynchronous system it is impossible to make sure that replicas of \emph{all} stale configurations
remove their private keys as it would require solving consensus~\cite{flp}.
However, as we show in this paper, it is possible to make sure that the configurations in which replicas do not
remove their keys are never accessed by correct clients and are incapable of creating ``proofs'' for output values.


Finally, there is a subtle, and quite interesting \textbf{``slow reader'' attack}.
Suppose that a client hears from \emph{almost all} replicas in a quorum of the current configuration
each holding a stale state, and not yet from the only correct replica in the quorum that has the up-to-date state.
The client then falls asleep.
Meanwhile, the configuration is superseded by a new one.
As we do not make any assumptions about the correctness of replicas in stale configurations,
the replica that has not yet responded can be compromised.
Moreover, due to asynchrony, this replica can still retain its original private key.
The replica can then pretend to be unaware of the current state.
Therefore, the slow client might still be able to complete its request in the superseded configuration
and return a stale state, which would violate the safety properties of the system.
\needsrev{%
In Section~\ref{sec:dbla-main}, we give a detailed example of this attack and show that
it can be addressed by an additional, ``confirming'' round-trip executed by the client.}

\myparagraph{Our contribution: Byzantine fault-tolerant reconfigurable services.}
We provide a systematic solution to each of the challenges described above
and present a set of techniques for building reconfigurable services in asynchronous model
with Byzantine faults of both clients and replicas.
We consider a very strong adversary: any number of clients can be Byzantine and,
as soon as some configuration is installed,
no assumptions are made about the correctness of replicas in any of the prior configurations.

Moreover, in our quest for a simple solution for the Byzantine model,
we devised a new approach to building asynchronous reconfigurable services
by further exploring the connection between reconfiguration and lattice
agreement~\cite{smartmerge, rla}.
We believe that this approach can be effectively applied to crash fault-tolerant systems as well.
\atadd{As we discuss in Section~\ref{subsec:transform-discussion}, the proposed protocol has the time complexity that is optimal even for crash fault-tolerant systems.}

Instead of trying to build a complex graph of configurations ``on the fly''
while simultaneously transferring the state between these configurations,
we start by simply assuming that we are already given a \emph{linear history}
(i.e., a sequence of configurations).
We introduce the notion of a \emph{{\dynamic} object} -- an object
that can transfer its own state between the configurations of a given finite linear history
and serve meaningful user requests.
We then provide {\dynamic} implementations of several important object types, such as
{\LatticeAgreement} and {\MaxRegister}. 
We expect that other asynchronous static algorithms
can be translated to the {\dynamic} model using similar techniques.

Finally, we present a \emph{general transformation} that allows us to combine \emph{any} {\dynamic} object
with two \emph{\DynamicByzantineLatticeAgreement} objects in such a way that together
they constitute a single \emph{\reconfigurable} object, which
exports a general-purpose reconfiguration interface and supports all the operations of the original
{\dynamic} object.

\atrev{This paper is a revised and extended version of a conference article~\cite{bla-disc}.}

\myparagraph{Roadmap.}
The rest of the paper is organized as follows.
We overview the model assumptions in Section~\ref{sec:system-model} and
define our principal abstractions in Section~\ref{sec:definitions}.
\atrev{In Section~\ref{sec:dbla-main}, we describe our implementation of {\DynamicByzantineLatticeAgreement}.
In Section~\ref{sec:transform-main}, we show how to use it to implement \atrev{reconfigurable objects} out of {\dynamic} ones.}
In Section~\ref{sec:access-control}, several possible implementations of access control are discussed.
We discuss related work in Section~\ref{sec:related-work} and conclude in Section~\ref{sec:discussion}.

\atrev{The proof of correctness for our {\DynamicByzantineLatticeAgreement} abstraction is delegated to Appendix~\ref{sec:dbla-correctness-proof}.
Finally, as an application of our constructions, we provide an implementation of a {\dynamic} {\MaxRegister} in Appendix~\ref{sec:max-register}.}



\section{System Model}\label{sec:system-model}

\myparagraph{Processes and channels.}
\oldconfirm{We consider a system of \emph{processes}.
A process can be a \emph{replica} or a \emph{client}.
Let $\Replicas$ and $\Clients$ denote the (possibly infinite) sets of replicas and clients, resp.,
that potentially can take part in the computation.}
At any point in a given execution, a process can be in one of the four states:
\emph{idle}, \emph{correct}, \emph{halted}, or \emph{Byzantine}.
\needsrev{Initially, each process is \emph{idle}.
An idle process does not participate in the protocol.
Once a process starts executing the protocol and as long as it does not execute the ``halt'' command and does not deviate from the prescribed protocol,
it is considered \emph{correct}.
A process is \emph{halted} if it executed the special
``halt'' command and stopped taking further steps.
Finally, a process is \emph{Byzantine} if it prematurely stops taking steps of the
algorithm or takes steps that are not prescribed by it.}
A correct process can later halt or become Byzantine.
However, the reverse is impossible: a halted or Byzantine process cannot become correct.
We assume that a process that remains correct forever (we call \needsrev{such processes} \emph{forever-correct})
does not prematurely stop taking steps of its algorithm.

We assume asynchronous \emph{reliable} \emph{authenticated} point-to-point
links between each pair of processes~\cite{cachin2011introduction}.
If a forever-correct process $p$ sends a message $m$ to a forever-correct process $q$,
then $q$ eventually  delivers $m$.
Moreover,  if a correct process $q$ receives a message $m$ from a process $p$ at time $t$,
and $p$ is correct at time $t$, then $p$ has indeed sent $m$ to $q$ before $t$.
%

We assume that the adversary is computationally bounded so that it is unable to break the cryptographic techniques,
such as digital signatures, forward security schemes~\cite{bellare1999forward} and one-way hash functions.

\myparagraph{Configuration lattice.}
A \emph{join semi-lattice} \oldconfirm{(or simply a \emph{lattice})} is a tuple  $(\Lat, \sqsubseteq)$, where $\Lat$ is a set
partially ordered by the binary relation $\sqsubseteq$ such that for
all elements $x,y \in \Lat$, there exists the \emph{least upper bound} for the set $\{x, y\}$,
i.e., the element $z \in \Lat$ such that $x, y \sqsubseteq z$
and $\forall\ w \in \Lat:$ if $x, y \sqsubseteq w$, then $z \sqsubseteq w$.
The least upper bound for the set $\{x, y\}$ is denoted by $x \sqcup y$.
$\sqcup$ is called the \emph{join operator}.
It is an associative, commutative, and idempotent binary operator on $\Lat$.
We write $x \sqsubset y$ whenever $x \sqsubseteq y$ and $x \neq y$.
We say that $x, y \in \Lat$ are \emph{comparable} iff either $x \sqsubseteq y$ or $y \sqsubset x$.

\oldconfirm{For any (potentially infinite) set $A$, $(2^A, \sqsubseteq)$ is a join semi-lattice,
called \emph{the powerset lattice of $A$}.
For all $Z_1, Z_2 \in 2^A$, $Z_1 \sqsubseteq Z_2 \bydef Z_1 \subseteq Z_2$ and $Z_1 \sqcup Z_2 \bydef Z_1 \cup Z_2$.}

A configuration is an element of a join semi-lattice $(\Conf, \sqsubseteq)$.
We assume that every configuration is associated with a finite set of
replicas via a map $\ireplicas: \Conf \to 2^{\Replicas}$,
and a \emph{quorum system} via a map $\iquorums: \Conf \to
2^{2^{\Replicas}}$, such that $\forall C \in \Conf: \iquorums(C)
\subseteq 2^{\ireplicas(C)}$.
Additionally we assume that there is a map $\iheight: \Conf \to \mathbb{Z}$, such that
$\forall C \in \Conf: \iheight(C) \ge 0$ and
$\forall C_1, C_2 \in \Conf:$ if $C_1 \sqsubset C_2$, then $\iheight(C_1) < \iheight(C_2)$.
%
We say that a configuration $C$ is \emph{\thigher} (resp., \emph{lower}) than a configuration $D$ iff $D \sqsubset C$ (resp, $C \sqsubset D$).
\needsrev{Note that ``$C$ is {\thigher} than $D$'' implies ``$\iheight(C) > \iheight(D)$'', but not vice versa.}

We say that $\iquorums(C)$ is a  \emph{dissemination quorum system}
at time $t$
iff every two sets (also called \emph{quorums}) in $\iquorums(C)$ have \oldconfirm{at least one replica in common
that is correct at time $t$},
and at least one quorum is  \emph{available} (all its replicas are correct) \oldconfirm{at time $t$}.
%
%

A natural (but not the only possible) way to define the lattice~$\Conf$ is as follows:
let $\Updates$ be $\{+,-\} \times \Replicas$, where
tuple $(+, p)$ means ``add replica $p$'' and tuple $(-, p)$ means ``remove replica $p$''.
Then $\Conf$ is the powerset lattice $(2^{\Updates}, \sqsubseteq)$.
The mappings $\ireplicas$, $\iquorums$, and $\iheight$ are defined as follows:
$\ireplicas(C) \bydef \{s \in \Replicas \mid (+, s) \in C \land (-, s) \notin C\}$,
%
It is straightforward to verify that $\iquorums(C)$
is a dissemination quorum system when strictly less than one third of replicas
in $\ireplicas(C)$ are faulty.
%
%
Note that, when this lattice is used for configurations,
once a replica is removed from the system, it cannot be added again with the same identifier.
In order to add such a replica back to the system, a new identifier must be used.
%

\myparagraph{Forward-secure digital signatures.}
%
%
%
In a \emph{forward-secure digital signature scheme}~\cite{bellare1999forward,malkin2002efficient,boyen2006forward,drijvers2019pixel},
the public key of a process is fixed while the secret key can evolve.
Each signature is associated with a \emph{timestamp}.
To generate a signature with timestamp $t$, the signer uses secret key $\isk_t$.
The signer can \emph{update its secret key} and get $\isk_{t_2}$ from $\isk_{t_1}$
if $t_1 < t_2 \le T$.\footnote{$T$ is a parameter of the scheme and can be set arbitrarily large
(with a modest overhead). We believe that $T = 2^{32}$ or $T = 2^{64}$ should be sufficient for most applications.}
However ``downgrading'' the key to a lower timestamp is computationally infeasible.
\needsrev{Thus}, if the signer updates their secret key to some timestamp $t$ and then removes the original secret key,
it will not be able to sign new messages with a timestamp lower than $t$, even if it later turns Byzantine.


For simplicity, we model a forward-secure signature scheme as an oracle
which associates every process $p$ with a timestamp $\ist_p$ (initially, $\ist_p = 0$).
The oracle provides $p$ with three operations:
(1) $\fUpdateFSKeys(t)$ sets $\ist_p$ to $t \ge \ist_p$;
(2) $\fFSSign(m, t)$ returns a signature for message $m$ and timestamp $t$ if $t \ge \ist_p$,
otherwise it returns $\bot$;
and (3) $\fFSVerify(m, p, s, t)$ returns $\iTrue$ iff $s$ was generated by invoking $\fFSSign(m, t)$
by process $p$.%
\footnote{We assume that anyone who knows the id of a process also knows its public key.
For example, the public key can be directly embedded into the identifier.}
In our protocols, we use the height of the configuration as the timestamp.
When a replica answers requests in configuration $C$,
it signs messages with timestamp $\iheight(C)$.
When a higher configuration $D$ is installed,
the replica invokes $\fUpdateFSKeys(\iheight(D))$.
This prevents the {\attackName} attack described in the introduction. 




\section{Abstractions and Definitions}\label{sec:definitions}

In this section, we introduce principal abstractions of this paper
(the access-control interface, {\ByzantineLatticeAgreement}, {\Reconfigurable} and {\Dynamic} objects),
state our quorum assumptions, and recall the definitions of broadcast primitives used in our algorithms.

\subsection{Access control \needsrev{and object composition}}

In our implementations and definitions, we parameterize some abstractions
by boolean functions $\fVerifyInputValue(v, \sigma)$ and $\fVerifyInputConfig(C, \sigma)$,
where $\sigma$ is called a \emph{certificate}.
Moreover, some objects also export a boolean function $\fVerifyOutputValue(v, \sigma)$,
which lets anyone to verify that the value $v$ was indeed produced by the object.
This helps us to deal with Byzantine clients.
In particular, it achieves three important goals.

First, the parameter $\fVerifyInputConfig$
allows us to prevent Byzantine clients from reconfiguring the system in an undesirable way
or flooding the system with excessively frequent reconfiguration requests.
In Section~\ref{sec:access-control}, we propose three simple implementations of this functionality:
each reconfiguration request must be signed by a quorum of replicas of some configuration\footnote{%
Additional care is needed to prevent the ``slow reader'' attack. See Section~\ref{sec:access-control} for more details.}
or by a quorum of preconfigured administrators.

Second, the parameter $\fVerifyInputValue(v, \sigma)$
allows us to formally capture the application-specific notions of well-formed client requests and access control.
For example, in a key-value storage system, each client can be permitted to modify only
the key-value pairs that were created by this client.
In this case, the certificate $\sigma$ is just a digital signature of the client.

Finally, the exported function $\fVerifyOutputValue$ allows us to compose several distributed objects
in such a way that the output of one object is passed as input for another one.
\needsrev{%
For example, in Section~\ref{sec:transform-main}, one object ($\iHistLA$) operates exclusively on outputs of
another object ($\iConfLA$).
We use the parameter function $\fVerifyInputValue$ of $\iHistLA$ and the exported function
$\fVerifyOutputValue$ of $\iConfLA$ to guarantee that a Byzantine client cannot
send to $\iHistLA$ a value that was not produced by $\iConfLA$.}

\subsection{{\ByzantineLatticeAgreement} abstraction}\label{subsec:bla-def}

In this section we formally define \emph{\ByzantineLatticeAgreement} abstraction ({\BLA} for short),
which serves as one of the main building blocks for constructing reconfigurable objects.
%
%
{\ByzantineLatticeAgreement} is an adaptation of {\LatticeAgreement}~\cite{gla}
that can tolerate Byzantine failures of processes (both clients and replicas).
It is parameterized by a join semi-lattice $\Lat$, called the \emph{object lattice},
and a boolean function $\fVerifyInputValue: \Lat \times \Sigma \to \{\iTrue, \iFalse\}$,
where $\Sigma$ is a set of possible certificates.
We say that $\sigma$ is a \emph{valid certificate for input value} $v$
iff $\fVerifyInputValue(v, \sigma) = \iTrue$.

We say that $v \in \Lat$ is a \emph{verifiable input value} in a given run iff at some point in time in that run, some process \emph{knows}
a certificate $\sigma$ that is valid for $v$, i.e., it maintains $v$ and a valid certificate $\sigma$ in its local memory.
We require that the adversary is unable to invert $\fVerifyInputValue$ by computing a valid certificate for a given value.
This is the case, for example, when $\sigma$  must contain a set of unforgeable digital signatures.
%

The {\ByzantineLatticeAgreement} abstraction exports one operation and one function.\footnote{%
  Recall that, unlike an operation,  a function can be computed locally,
  without communicating with other processes, and the result only depends on the function's input.}
\begin{itemize}
  \item Operation $\fPropose(v, \sigma)$ returns a response of the form $\Tupple{w, \tau}$,
    where $v, w \in \Lat$, $\sigma$ is a valid certificate for input value $v$,
    and $\tau$ is a certificate for output value $w$;

  \item Function $\fVerifyOutputValue(v, \sigma)$ returns a boolean value.
\end{itemize}
Similarly to input values, we say that $\tau$ is a \emph{valid certificate for output value} $w$
iff $\fVerifyOutputValue(w, \tau) = \iTrue$.
We say that $w$ is a \emph{verifiable output value} in a given run iff at some point in that run, some process knows
$\tau$ that is valid for $w$.

Implementations of {\ByzantineLatticeAgreement} must satisfy the following properties:
\begin{itemize}
  \item \textit{\pBLAValidity}:
    Every verifiable output value $w$ is a join of some set of verifiable input values;

  \item \textit{\pBLAVerifiability}:
    If $\fPropose(\ldots)$ returns $\Tupple{w, \tau}$ to a correct process,
    then $\fVerifyOutputValue(w, \tau) = \iTrue$;

  \item \textit{\pBLAInclusion}:
    If $\fPropose(v, \sigma)$ returns $\Tupple{w, \tau}$ to a correct process, then $v \sqsubseteq w$;

  \item \textit{\pBLAComparability}:
    All verifiable output values are comparable;


 \item \textit{\pBLALiveness}:
    If the total number of verifiable input values is finite,
    every call to $\fPropose(v, \sigma)$ by a forever-correct process eventually returns.
\end{itemize}
For the sake of simplicity, we only guarantee liveness when there are finitely many verifiable input values.
This is sufficient for the purposes of reconfiguration, as it only guarantees liveness under the assumption that only finitely many valid  reconfiguration calls are issued.
In practice, this assumption boils down to providing liveness when not ``too many'' conflicting values are concurrently proposed.
The abstraction that provides unconditional liveness is called \emph{\GeneralizedLatticeAgreement}~\cite{gla}.

\subsection{{\Reconfigurable} objects}\label{subsec:reconf-objects}

It is possible to define a \emph{\reconfigurable} version of every static distributed object
by enriching its interface and imposing some additional properties.
In this section, we define the notion of a {\reconfigurable} object in a very abstract way.
By combining this definition with the definition of a {\ByzantineLatticeAgreement} from Section~\ref{subsec:bla-def},
we obtain a formal definition of a {\ReconfigurableByzantineLatticeAgreement}.
Similar combination can be performed with the definition of any static distributed object
(e.g., with the definition of a {\MaxRegister} from Appendix~\ref{sec:max-register}).


\oldconfirm{%
A {\reconfigurable} object exports
an operation $\fUpdateConfig(C, \sigma)$, which can be used to reconfigure the system,}
and must be parameterized by a boolean function
$\fVerifyInputConfig: \Conf \times \Sigma \to \{\iTrue, \iFalse\}$,
where $\Sigma$ is a set of possible certificates.
\needsrev{Similarly to verifiable input values,}
we say that $C \in \Conf$ is a \emph{verifiable input configuration} in a given run iff
at some point in that run, some process knows $\sigma$
such that $\fVerifyInputConfig(C, \sigma) = \iTrue$.
%
%
%
%

We require the total number of verifiable input configurations to be finite
in any given infinite execution of the protocol.
%
%
In practice, this boils down to assuming sufficiently long periods of stability
when no new verifiable input configurations appear.
This requirement is imposed by all asynchronous reconfigurable
storage systems~\cite{tutorial2010,SKM17-reconf,rla,freestore}
we are aware of,
and, in fact, can be shown to be necessary~\cite{r-liveness}.



\oldconfirm{%
When a correct replica $r$ is ready to serve user requests in a configuration $C$,
it triggers upcall $\fInstalledConfig(C)$.
We then say that $r$ \emph{installs} configuration $C$.}
\needsrev{At any given moment in time,
a configuration is called \emph{installed} if some correct replica has installed it
and it is called \emph{\tsuperseded} if some {\thigher} configuration is installed.}


\oldconfirm{Each {\reconfigurable} object must satisfy the following properties:
\begin{itemize}
  \item \textit{\pReconfigurationValidity}:
    Every installed configuration $C$ is a join of some set of verifiable input configurations.
    Moreover, all installed configurations are comparable;

  \item \oldconfirm{\textit{\pReconfigurationLiveness}:
    Every call to $\fUpdateConfig(C, \sigma)$ by a forever-correct client eventually returns.
    Moreover, $C$ or a {\thigher} configuration will eventually be installed.}

  \item \textit{\pInstallationLiveness}:
    %
    If some configuration $C$ is installed by some correct replica,
    then $C$ or a {\thigher} configuration will eventually be installed by all correct replicas.
\end{itemize}}
%


\subsection{{\Dynamic} objects}\label{subsec:dynamic-objects}

{\Reconfigurable} objects are hard to build because they need to solve two problems at once.
First, they need to order and combine concurrent reconfiguration requests.
Second, the state of the object needs to be transferred across installed configurations.
We decouple these two problems by introducing the notion of a \emph{\dynamic} object.
{\Dynamic} objects solve the second problem while ``outsourcing'' the first one.

Before we formally define {\dynamic} objects, let us first define the notion of a \emph{history}.
In Section~\ref{sec:system-model}, we introduced the configuration lattice $\Conf$.
A finite set $h \subseteq \Conf$ is called a \emph{history} iff all elements of $h$ are comparable
(in other words, if they form a sequence).
Let $\fMaxElement(h)$ be $C \in h$ such that $\forall\ C' \in h: C' \sqsubseteq C$.
\pkrev{$\fMaxElement(h)$ is well-defined, as the configurations in $h$ are totally ordered.}

{\Dynamic} objects must export an operation $\fUpdateHistory(h, \sigma)$
and must be parameterized by a boolean function $\fVerifyHistory: \Hist \times \Sigma \to \{\iTrue, \iFalse\}$,
where $\Hist$ is the set of all histories and $\Sigma$ is the set of all possible certificates.
We say that $h$ is a \emph{verifiable history} in a given run iff at some point in that run, some process knows $\sigma$ such that $\fVerifyHistory(h, \sigma) = \iTrue$.
A configuration $C$ is called \emph{candidate} iff it belongs to some verifiable history.
Also, a candidate configuration $C$ is called \emph{\tactive} iff it is not {\tsuperseded} by a {\thigher} configuration.

As with verifiable input configurations,
the total number of verifiable histories is required to be finite.
Additionally, we require all verifiable histories to be related by containment (i.e., comparable w.r.t. $\subseteq$).
\pkrev{Recall that a history is a totally ordered (w.r.t. $\sqsubseteq$) set of configurations.}
Formally, if $\fVerifyHistory(h_1, \sigma_1) = \iTrue$ and $\fVerifyHistory(h_2, \sigma_2) = \iTrue$,
then $h_1 \subseteq h_2$ or $h_2 \subseteq h_1$.
We discuss how to build such histories in Section~\ref{sec:transform-main}.

Similarly to {\reconfigurable} objects, a {\dynamic} object must have the $\fInstalledConfig(C)$ upcall.
The object must satisfy the following properties:
\begin{itemize}
  \item \textit{\pDynamicValidity}:
    Only a candidate configuration can be installed by a correct replica;%

  \item \textit{\pDynamicLiveness}:
    Every call to $\fUpdateHistory(h, \sigma)$ by a forever-correct client eventually returns.
    Moreover, $\fMaxElement(h)$ or a {\thigher} configuration will eventually be installed;

  \item \textit{\pInstallationLiveness} (the same as for {\reconfigurable} objects):
    If some configuration $C$ is installed by some correct replica,
    then $C$ or a {\thigher} configuration will eventually be installed by all correct replicas.
\end{itemize}
Note that {\pDynamicValidity} implies that all installed configurations are comparable,
since all verifiable histories are related by containment and all configurations within one history are
comparable.

While {\reconfigurable} objects provide general-purpose reconfiguration interface,
{\dynamic} objects are weaker,
as they require an external service to build comparable verifiable histories.
%
As the main contribution of this paper,
we show how to build \emph{\dynamic} objects in a Byzantine environment
and how to create \emph{\reconfigurable} objects using
{\dynamic} objects as building blocks.
We argue that this technique is applicable to a large class of objects.

\subsection{Quorum system assumptions}\label{subsec:intersection-and-availability}

Most fault-tolerant implementations of distributed objects impose some requirements on the subsets of
processes that can be faulty.
We say that a configuration $C$ is \needsrev{\emph{correct at time $t$}}
iff $\ireplicas(C)$ is a dissemination quorum system at time $t$ (as defined in Section~\ref{sec:system-model}).
Correctness of our implementations of \emph{\dynamic} objects
relies on the assumption that {\tactive} candidate configurations are correct.
Once a configuration is superseded by a {\thigher} configuration, we make no further assumptions about it.

For \emph{\reconfigurable} objects we impose a slightly more conservative requirement:
every combination of verifiable input configurations that is not yet superseded must be correct.
Formally, we require:
\begin{description}
\item[Quorum availability:]  Let $C_1, \dots, C_k$ be verifiable input configurations such that $C = C_1 \sqcup \dots \sqcup C_k$
is not superseded at time $t$.
Then we require $\iquorums(C)$ to be a dissemination quorum system at time $t$.
\end{description}

Correctness of our \emph{\reconfigurable} objects relies solely on correctness of the {\dynamic} building blocks.
Formally, when $k$ configurations are concurrently proposed, we require all possible combinations,
i.e., $2^k - 1$ configurations, to be correct.
However, in practice, at most $k$ of them will be chosen to be put in verifiable histories,
and only those configurations will be accessed by correct processes.
We impose a more conservative requirement because we do not know these configurations \emph{a priori}.

\subsection{Broadcast primitives}\label{subsec:bcast}

To make sure that no process is ``left behind'',
we assume that a variant of \emph{reliable broadcast primitive}~\cite{cachin2011introduction} is available.
The primitive must ensure two properties:
\begin{enumerate}
\item[(1)] If a forever-correct process~$p$ broadcasts a message~$m$, then $p$ eventually delivers $m$;
\item[(2)] If some message~$m$ is delivered by a \emph{forever-correct} process, every forever-correct process eventually delivers $m$.
\end{enumerate}
%
%
\needsrev{Note that we do not make any assumptions involving any processes that are not forever-correct.}
\needsrev{In practice such a primitive can be implemented by a gossip protocol~\cite{gossiping}.}
This primitive is ``global'' in a sense that it is not bound to any particular configuration.
In pseudocode we use ``{\RBBroadcast{\ldots}}'' to denote a call to the ``global'' reliable broadcast.
%

Additionally, we assume a ``local'' \emph{uniform reliable broadcast} primitive~\cite{cachin2011introduction}.
It has a stronger totality property: if some \emph{correct} process $p$ delivered some message $m$,
then every forever-correct process will eventually deliver $m$, even if $p$ later turns Byzantine.
This primitive can be implemented in a \emph{static} system, provided a quorum system.
As we deal with \emph{\dynamic} systems, we associate every broadcast message with a fixed configuration
and only guarantee these properties if the configuration is \emph{never superseded}.
Note that any static implementation of uniform reliable broadcast trivially guarantees this property.
In pseudocode we use ``{\URBBroadcastIn{\ldots}{$C$}}'' to denote a call to the ``local'' uniform reliable broadcast
in configuration $C$.


\section{\DynamicByzantineLatticeAgreement}\label{sec:dbla-main}

\begin{algorithm*}
    \caption{DBLA object specification}
    \labelalg{dbla-spec}
    
    \BeginAlgorithmic
        \Parameters
            \State Lattice of configurations $\Conf$ and the initial configuration $\Cinit \in \Conf$
            \State The object lattice $\Lat$ and the initial value $\Vinit \in \Lat$
            \State Boolean functions $\fVerifyHistory(h, \sigma)$ and $\fVerifyInputValue(v, \sigma)$
        \EndParameters
        
        \algspace
        \Interface
            \Operation{$\fPropose$}{$v$, $\sigma$} \EndOperation
            \Operation{$\fUpdateHistory$}{$h$, $\sigma$} \EndOperation
            \Function{$\fVerifyOutputValue$}{$v$, $\sigma$} \EndFunction
            \Upcall{$\fInstalledConfig$}{$C$} \EndUpcall
        \EndInterface
        
        \algspace
        \PropertiesInline
                {\pBLAValidity},
                {\pBLAVerifiability},
                {\pBLAInclusion},
                {\pBLAComparability},
                {\pBLALiveness},
                \\ \hspace{\algorithmicindent}
                {\pDynamicValidity},
                {\pDynamicLiveness},
                {\pInstallationLiveness}
    \EndAlgorithmic
\end{algorithm*}

\emph{\DynamicByzantineLatticeAgreement} abstraction ({\DBLA} for short) is the main building block
in our construction of {\reconfigurable} objects.
Its specification is a combination of the specification of {\ByzantineLatticeAgreement} (Section~\ref{subsec:bla-def})
and the specification of a {\dynamic} object (Section~\ref{subsec:dynamic-objects}).
\atadd{It is summarized in Algorithm~\lstref{dbla-spec}.}
%
%
\atrev{In this section, we provide the implementation of {\DBLA} and analyze the time complexity of the solution.
The proof of correctness is delegated to Appendix~\ref{sec:dbla-correctness-proof}.}

As we mentioned earlier, we use forward-secure digital signatures to guarantee that superseded configurations
cannot affect correct clients or forge certificates for output values.
Ideally, before a new configuration $C$ is installed
(i.e., before a correct replica triggers $\fInstalledConfig(C)$ upcall),
we would like to make sure that the replicas of all candidate configurations {\tlower}
than $C$ invoke $\fUpdateFSKeys(\iheight(C))$.
However, this would require the replica to know the set of all candidate configurations {\tlower} than $C$.
Unambiguously agreeing on this set would require solving consensus,
which is known to be impossible in a fault-prone asynchronous system~\cite{flp}.

Instead, we classify all candidate configurations in two categories:
\emph{\pivotal} and \emph{\tentative}.
A candidate configuration is called \emph{\pivotal} if it is the {\thighest} configuration in some verifiable history.
Otherwise it is called \emph{\tentative}.
A nice property of {\pivotal} configurations is that it is impossible to ``skip'' one in a verifiable history.
Indeed, if $C_1 = \fMaxElement(h_1)$ and $C_2 = \fMaxElement(h_2)$ and $C_1 \sqsubset C_2$,
then, since all verifiable histories are related by containment, $h_1 \subseteq h_2$ and $C_1 \in h_2$.
This allows us to make sure that, before a configuration $C$ is installed,
the replicas in all {\pivotal} (and, possibly, some {\tentative})
configurations {\tlower} than $C$ update their keys.

In order to reconfigure a {\DBLA} object, a correct client must use reliable broadcast
to distribute the new verifiable history.
Each correct process $p$ maintains, locally, the largest (with respect to $\subseteq$) verifiable history it
delivered so far through reliable broadcast.
It is called \emph{the local history of process $p$} and is denoted by $\ihistory_p$.
We use $\Chighest_p$ to denote the most recent configuration in $p$'s local history
(i.e., $\Chighest_p = \fMaxElement(\ihistory_p)$).
Whenever a replica $r$ updates $\ihistory_r$, it invokes $\fUpdateFSKeys(\iheight(\Chighest_r))$.
Recall that if at least one forever-correct process delivers a message via reliable broadcast,
every other forever-correct process will eventually deliver it as well.


Similarly, each process $p$ keeps track of all verifiable input values it has seen
$\icurVals_p \subseteq \Lat \times \Sigma$, where $\Lat$ is the object lattice and $\Sigma$ is the set of all possible certificates.
Sometimes, during the execution of the protocol, processes exchange these sets.
Whenever a process $p$ receives a message that contains a set of values with certificates $\ivalues \subseteq \Lat \times \Sigma$,
it checks that the certificates are valid (i.e., $\forall\ (v, \sigma) \in \ivalues: \fVerifyInputValue(v, \sigma) = \iTrue$)
and adds these values and certificates to $\icurVals_p$.


\subsection{Client implementation}

\begin{algorithm*}
    \caption{{\DBLA}: code for client $p$ (part 1)}
    \labelalgpart{dbla-client}{1}


  \BeginAlgorithmic
    \Parameters
      \State Lattice of configurations $\Conf$ and the initial configuration $\Cinit$
      \State The object lattice $\Lat$ and the initial value $\Vinit$
      \State Boolean functions $\fVerifyHistory(h, \sigma)$ and $\fVerifyInputValue(v, \sigma)$
    \EndParameters

    \Globals
      \State $\ihistory \subseteq \Conf$, initially $\{ \Cinit \}$                                        \Comment{local history of this process}
      \State $\Sigmahistory \in \Sigma$, initially $\bot$                                                 \Comment{proof for the local history}
      \State $\icurrentValues \subseteq \Lat \times \Sigma$, initially $\{ \Tupple{\Vinit, \bot} \}$      \Comment{known verifiable input values with proofs}
      \State $\istatus \in \{ \iinactive, \iproposing, \iconfirming \}$, initially $\iinactive$
      \State $\iseqNumber \in \mathbb{Z}$, initially $0$        \Comment{used to match requests with responses}
      \State $\iacks_1$, initially $\emptyset$                  \Comment{a set of pairs of form $\Tupple{\iprocessId, \isignature}$}
      \State $\iacks_2$, initially $\emptyset$                  \Comment{a set of pairs of form $\Tupple{\iprocessId, \isignature}$}
    \EndGlobals

    \AuxiliaryFunctions
      \State $\fMaxElement(h)$ \Comment{returns the {\thighest} configuration in history $h$}
      \State $\fContainsQuorum(\iacks, C)$ \Comment{returns $\iTrue$ iff ${\exists} Q \in \iquorums(C) \suchThat {\forall} r \in Q: \Tupple{r, *} \in \iacks$}
      \State $\fJoinAll(\ivs)$ \Comment{returns the lattice join of all elements in $\ivs$}
      \State $\fVerifyValues(\ivs)$ \Comment{returns $\iTrue$ iff $\forall \Tupple{v, \sigma} \in \ivs: \fVerifyInputValue(v, \sigma)$}
      \State $\fFSVerify(m, r, s, t)$ \Comment{verifies forward-secure signature (see Section~\ref{sec:system-model})}
    \EndAuxiliaryFunctions

    \algspace
    \Operation{$\fPropose$}{$v$, $\sigma$} \labelline{propose}
      \State $\fRefine(\{ \Tupple{v, \sigma} \})$ \labelline{propose-refine}
      \State \WaitFor $\fContainsQuorum(\iacks_2, \fMaxElement(\ihistory))$ \labelline{propose-wait}
      \State $\istatus \gets \iinactive$ \labelline{propose-after-wait}
      \State let $\sigma = \Tupple{\icurrentValues, \ihistory, \Sigmahistory, \iacks_1, \iacks_2}$                      \labelline{propose-make-proof}
      \State \Return $\Tupple{\fJoinAll(\icurrentValues), \sigma}$                                                      \labelline{propose-return}
    \EndOperation

    \algspace
    \Operation{$\fUpdateHistory$}{$h$, $\sigma$}                                                                        \labelline{update-history}
      \State \RBBroadcast{$\mNewHistory$, $h$, $\sigma$}
    \EndOperation

    \algspace    
    \Function{$\fVerifyOutputValue$}{$v$, $\sigma$} \labelline{verify-output-value}
      \If {$\sigma = \bot$} \Return $v = \Vinit$ \EndIf
      \State let $\Tupple{\ivalues, h, \sigma_h, \iproposeAcks, \iconfirmAcks} = \sigma$ \labelline{verify-output-value-unpack}
      \State let $C = \fMaxElement(h)$
      \State \Return $\fJoinAll(\ivalues) = v \land \fVerifyHistory(h, \sigma_h)$ \labelline{verify-output-value-verify-history}
          $\tlaStyleLand \fContainsQuorum(\iproposeAcks, C) \land \fContainsQuorum(\iconfirmAcks, C)$
          $\tlaStyleLand \forall\; \Tupple{r, s} \in \iproposeAcks: \fFSVerify(\Tupple{\mProposeResp, \ivalues}, r, s, \iheight(C))$
          $\tlaStyleLand \forall\; \Tupple{r, s} \in \iconfirmAcks: \fFSVerify(\Tupple{\mConfirmResp, \iproposeAcks}, r, s, \iheight(C))$ \labelline{verify-output-value-end}
    \EndFunction
    
  \EndAlgorithmic
\end{algorithm*}

\begin{algorithm*}
  \caption{{\DBLA}: code for client $p$ (part 2)}
  \labelalgpart{dbla-client}{2}
  
  \BeginAlgorithmic

    \algspace
    \Procedure{$\fRefine$}{$\ivalues$}                                                                                  \labelline{refine}
      \State $\iacks_1 \gets \emptyset$; $\iacks_2 \gets \emptyset$
      \State $\icurrentValues \gets \icurrentValues \cup \ivalues$                                                      \labelline{refine-current-values}
      \State $\iseqNumber \gets \iseqNumber + 1$
      \State $\istatus \gets \iproposing$
      \State let $C = \fMaxElement(\ihistory)$
      \State \Send{$\mPropose$, $\icurrentValues$, $\iseqNumber$, $C$}{$\ireplicas(C)$}                                 \labelline{refine-send-message}
                                                                                                                        \labelline{refine-end}
    \EndProcedure

    \algspace
    \Upon{$\fContainsQuorum(\iacks_1, \fMaxElement(\ihistory))$}                                                        \labelline{upon-acks-collected}
      \State $\istatus \gets \iconfirming$
      \State let $C = \fMaxElement(\ihistory)$
      \State \Send{$\mConfirm$, $\iacks_1$, $\iseqNumber$, $C$}{$\ireplicas(C)$}                                        \labelline{upon-acks-collected-send-confirm}
    \EndHandler

    \algspace
    \UponReceive{$\mProposeResp$, $\ivalues$, $\isignature$, $\isn$}{replica $r$}                                       \labelline{upon-propose-resp}
      \State let $\isigValid = \fFSVerify(\Tupple{\mProposeResp, \ivalues}, r, \isignature, \iheight(\fMaxElement(\ihistory)))$
      \If {$\istatus = \iproposing \land \isn = \iseqNumber \land \isigValid$} \labelline{upon-propose-resp-check}
        \If {$\ivalues \nsubseteq \icurrentValues \land \fVerifyValues(\ivalues \setminus \icurrentValues)$} $\fRefine(\ivalues)$ \labelline{upon-propose-resp-refine}
        \ElsIf {$\ivalues = \icurrentValues$} $\iacks_1 \gets \iacks_1 \cup \{ \Tupple{r, \isignature} \}$              \labelline{upon-propose-resp-ack}
        \EndIf
      \EndIf                                                                                                            \labelline{upon-propose-resp-end}
    \EndHandler

    \algspace
    \UponReceive{$\mConfirmResp$, $\isignature$, $\isn$}{replica $r$}
      \State let $\isigValid = \fFSVerify(\Tupple{\mConfirmResp, \iacks_1}, r, \isignature, \iheight(\fMaxElement(\ihistory)))$
      \If {$\istatus = \iconfirming \land \isn = \iseqNumber \land \isigValid$}
        $\iacks_2 \gets \iacks_2 \cup \{ \Tupple{r, \isignature} \}$
      \EndIf
    \EndHandler

    \algspace
    \UponRBDeliver{$\mNewHistory$, $h$, $\sigma$}{any sender}                                                           \labelline{new-history}
      \If {$\fVerifyHistory(h, \sigma) \land \ihistory \subset h$}                                                      \labelline{new-history-verify}
        \State $\ihistory \gets h$; $\Sigmahistory \gets \sigma$                                                        \labelline{history-update}
        \If {$\istatus \in \{ \iproposing, \iconfirming \}$} $\fRefine(\emptyset)$ \EndIf                               \labelline{new-history-refine}
      \EndIf
    \EndHandler
  \EndAlgorithmic
\end{algorithm*}

The client's protocol is simple.
\atadd{The pseudocode for it is presented in Algorithms~\lstref{dbla-client-part-1} and~\lstref{dbla-client-part-2}.
Note that we omit the subscript $p$ in the pseudocode because each process can access only its own variables directly.}

As we mentioned earlier, the operation $\fUpdateHistory(h, \sigma)$ is implemented as
{\RBBroadcast{$\mNewHistory, h, \sigma$}} \atadd{(\lstlineref{dbla-client}{update-history})}.
The rest of the reconfiguration process is handled by the replicas.
The protocol for the operation $\fPropose(v, \sigma)$ \atadd{(\lstlinerangeref{dbla-client}{propose}{propose-return})} consists of two stages:
\emph{proposing} a value and \emph{confirming} the result.

The first stage (proposing) mostly follows the
implementation of lattice agreement by Faleiro et al.~\cite{gla}.
Client $p$ repeatedly sends message \Message{$\mPropose$, $\icurVals_p$, $\iseqNumber_r$, $C$}
to all replicas in $\ireplicas(C)$,
where $\mPropose$ is the message descriptor,
$C = \Chighest_p$,
and $\iseqNumber_r$ is a sequence number used by the client to match sent messages with replies.

After sending these messages to $\ireplicas(C)$, the client waits for responses of
the form \Message{$\mProposeResp$, $\ivalues$, $\isignature$, $\isn$},
where $\mProposeResp$ is the message descriptor,
$\ivalues$ is the set of all verifiable input values known to the replica with valid certificates
(including those sent by the client),
$\isignature$ is a forward-secure signature with timestamp $\iheight(C)$,
and $\isn$ is the same sequence number as in the message from the client.

During the first stage, one of the following cases can take place:
(1)~the client learns about some new verifiable input values from one of the $\mProposeResp$ messages;
(2)~the client updates its local history (by delivering it through reliable broadcast); and
(3)~the client receives a quorum of valid replies with the same set of verifiable input values.
%
In the latter case, the client proceeds to the second stage.
In the first two cases, the client simply restarts the operation.
%
%
\atrev{%
Recall that, according to the {\pBLALiveness} property, termination of client requests is only guaranteed
when the number of verifiable input values is finite.
Additionally, the number of verifiable histories is assumed to be finite.
Hence, the number of restarts will also be finite.
This is the main intuition behind the liveness of the client's protocol.}

\begin{figure}
    \begin{executionexample}
        \centering
        \pdftex[\linewidth]{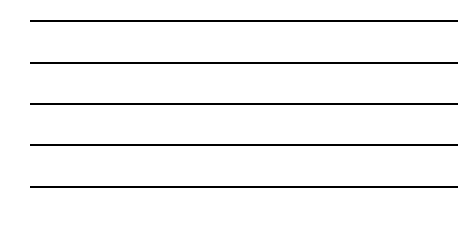}
        \caption{%
            \atrev{%
            An example execution of two concurrent $\fPropose$ operations.
            Solid black arrows and dashed blue arrows
            represent the messages exchanged in the first and the second
            stages of the $\fPropose$ protocol respectively.
            The sets of numbers represent the sets of verifiable input values known to the processes.
            Replica $r_3$ is Byzantine and always responds to $\mPropose$ messages with the
            same set of verifiable input values as in the incoming itself.}}
        \label{fig:two-concurrent-proposes}
    \end{executionexample}
\end{figure}

The example in Figure~\ref{fig:two-concurrent-proposes} illustrates how the first stage of the algorithm
ensures the comparability of the results when no reconfiguration is involved.
In this example, clients $p$ and $q$ concurrently propose values $\{1\}$ and $\{2\}$, respectively,
from the lattice $\Lat = 2^{\mathbb{N}}$.
Client $p$ successfully returns the proposed value $\{1\}$ while
client $q$ is forced to refine its proposal and return the combined value $\{1, 2\}$.
The quorum intersection prevents the clients from returning incomparable values (e.g., x$\{1\}$ and $\{2\}$).

In the second (confirming) stage of the protocol,
the client simply sends the acknowledgments it has collected in the first stage
to the replicas of the same configuration.
The client then waits for a quorum of replicas to reply with a forward-secure signature with
timestamp $\iheight(C)$.

\begin{figure}
    \centering
    \begin{executionexample}
        \pdftex[\linewidth]{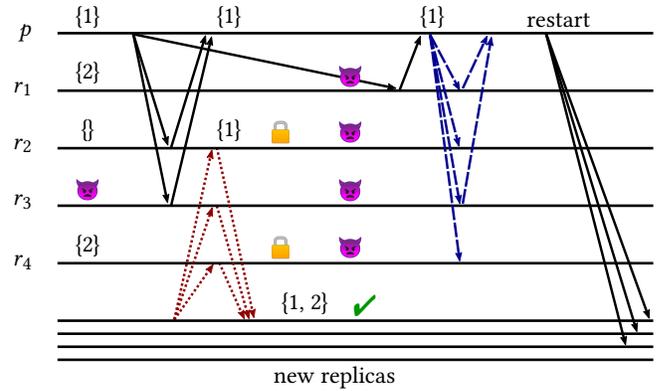}
        \def\svgwidth{\linewidth}
        \caption{%
            \atrev{%
            An example execution of a $\fPropose$ operation concurrent with a reconfiguration.
            The notation follows the convention from Figure~\ref{fig:two-concurrent-proposes}.
            Additionally, dotted red lines represent the messages exchanged during reconfiguration,
            the {\textlock} symbol means that the replica has updated its private key and can no longer serve client requests in this configuration, and the {\textevil} symbol means that the replica has become Byzantine.}}
        \label{fig:slow-reader-demonstation}
    \end{executionexample}
\end{figure}

\needsrev{%
The example in Figure~\ref{fig:slow-reader-demonstation} illustrates how reconfiguration can interfere
with an ongoing $\fPropose$ operation in what we call the \textbf{``slow reader'' attack},
and how the second stage of the protocol prevents a safety violation.
Imagine that a correct client completed the $\fPropose$ operation and received value $\{2\}$ before client $p$ started the execution.
As a result, all correct replicas in quorum $\{r_1, r_3, r_4\}$ store value $\{2\}$.
Then client $p$ executes $\fPropose(\{1\}, \sigma)$, where $\sigma$ is a valid certificate for input value $\{1\}$.
Due to the {\pBLAComparability}, {\pBLAInclusion}, and {\pBLAValidity} properties of {\ByzantineLatticeAgreement},
the only valid output value for client $p$ is $\{1,2\}$ (assuming that there are no other verifiable input values).
The client successfully reaches replicas $r_2$ and $r_3$ before the reconfiguration.
Neither $r_1$ nor $r_2$ tells the client about the input value $\{2\}$: $r_2$ is outdated and $r_3$ is Byzantine.
The message from $p$ to $r_1$ is delayed.
Meanwhile, a new configuration is installed, and all replicas of the original configuration become Byzantine.
When the message from $p$ finally reaches $r_1$,
the replica is already Byzantine and it can pretend that it has not seen any verifiable input values other than $\{1\}$.
The client then finishes the first stage of the protocol with value $\{1\}$.
Returning this value from from the $\fPropose$ operation would violate {\pBLAComparability}.

Luckily, the second stage of the protocol prevents the safety violation.
Since replicas $r_2$ and $r_4$ updated their private keys during the reconfiguration,
they are unable to send the signed confirmations with timestamp $\iheight(C)$ to the client.
Hence, the client will not be able to complete the operation in configuration $C$ and will wait until
it receives a new verifiable history via reliable broadcast and
will restart the operation in a {\thigher} configuration.}

The certificate for the output value $v \in \Lat$ produced by the $\fPropose$ protocol in a configuration $C$ consists of: 
(1)~the set of verifiable input values (with certificates for them) from the first stage of the algorithm
(the join of all these values must be equal to $v$);
(2)~\atrev{a verifiable history (with a certificate for it) that confirms that $C$ is {\pivotal} (i.e., for which $C$ is the {\thighest} configuration)};
(3)~the quorum of signatures from the first stage of the algorithm; and
(4)~the quorum of signatures from the second stage of the algorithm.
Intuitively, the only way for a Byzantine client to obtain such a certificate is to benignly follow the $\fPropose$ protocol.

\atrev{%
It is important that only {\pivotal} configurations can produce valid certificates because non-{\pivotal} ({\tentative}) configurations may contain fully compromised quorums with non-updated private keys for the forward-secure digital signature scheme.}

\subsection{Replica implementation}

\begin{algorithm*}
  \caption{{\DBLA}: code for replica $r$ (part 1)}
  \labelalgpart{dbla-replica}{1}

  \BeginAlgorithmic
    \ParametersInline $\Conf$, $\Lat$, $\Cinit$, $\Vinit$, $\fVerifyHistory(h, \sigma)$, and $\fVerifyInputValue(v, \sigma)$ (see Algorithm~\lstref{dbla-client-part-1})

    \Globals
      \State $\ihistory \subseteq \Conf$, initially $\{ \Cinit \}$                        \Comment{local history of this process}
      \State $\icurrentValues \subseteq \Lat \times \Sigma$, initially $\{ \Tupple{\Vinit, \bot} \}$    \Comment{known verifiable input values with proofs}
      \State $\Ccurr \in \Conf$, initially $\Cinit$                                       \Comment{current configuration}
      \State $\Cinst \in \Conf$, initially $\Cinit$                                       \Comment{installed configuration}
      \State $\iseqNumber \in \mathbb{Z}$, initially $0$                                  \Comment{used to match requests with responses}
      \State $\iactiveStateTransfer \in \{\iTrue, \iFalse\}$, initially $\iFalse$
    \EndGlobals

    \AuxiliaryFunctions
      \State $\fMaxElement$, $\fContainsQuorum$, $\fJoinAll$, $\fVerifyValues$ (see Algorithm~\lstref{dbla-client-part-1}).
      \State $\fFSSign(\mathit{message}, \mathit{timestamp})$ \Comment{produces a forward-secure signature (see Section~\ref{sec:system-model})}
      \State $\fUpdateFSKeys(t)$ \Comment{updates the signing timestamp (see Section~\ref{sec:system-model})}
    \EndAuxiliaryFunctions

    \algspace
    \UponReceive{$\mPropose$, $\ivalues$, $\isn$, $C$}{client $c$}                                                      \labelline{upon-propose}
      \State \WaitFor $C = \Cinst \lor \fMaxElement(\ihistory) \not\sqsubseteq C$                                       \labelline{upon-propose-wait}
      \If {$C = \fMaxElement(\ihistory) \land \fVerifyValues(\ivalues \setminus \icurrentValues)$}
        \State $\icurrentValues \gets \icurrentValues \cup \ivalues$                                                    \labelline{upon-propose-cur-vals}
        \State let $\isignature = \fFSSign(\Tupple{\mProposeResp, \icurrentValues}, \iheight(C))$                       \labelline{upon-propose-sign}
        \State \Send{$\mProposeResp$, $\icurrentValues$, $\isignature$, $\isn$}{$c$}
      \Else{} ignore the message
      \EndIf                                                                                                            \labelline{upon-propose-end}
    \EndHandler

    \algspace
    \UponReceive{$\mConfirm$, $\iproposeAcks$, $\isn$, $C$}{client $c$}                                                 \labelline{upon-confirm}
      \State \WaitFor $C = \Cinst \lor \fMaxElement(\ihistory) \not\sqsubseteq C$                                       \labelline{upon-confirm-wait}
      \If {$C = \fMaxElement(\ihistory)$}
        \State let $\isignature = \fFSSign(\Tupple{\mConfirmResp, \iproposeAcks}, \iheight(C))$                         \labelline{upon-confirm-sign}
        \State \Send{$\mConfirmResp$, $\isignature$, $\isn$}{$c$}
      \Else{} ignore the message
      \EndIf
    \EndHandler
  \EndAlgorithmic
\end{algorithm*}

\begin{algorithm*}
  \caption{{\DBLA}: code for replica $r$ (part 2)}
  \labelalgpart{dbla-replica}{2}

  \BeginAlgorithmic
    \LineCommentx{State transfer}
    \Upon{$\Ccurr \neq \fMaxElement(\{ C \in \ihistory \mid r \in \ireplicas(C) \}) \land \text{not } \iactiveStateTransfer$}                                    \labelline{state-transfer-begin}
      \State let $\Cnext = \fMaxElement(\{ C \in \ihistory \mid r \in \ireplicas(C) \})$                                 \labelline{state-transfer-first-line}
      \State let $S = \{ C \in \ihistory \mid \Ccurr \sqsubseteq C \sqsubset \Cnext \}$                                  \labelline{state-transfer-speculation}
      \State $\iactiveStateTransfer \gets \iTrue$
      \State $\iseqNumber \gets \iseqNumber + 1$
      \ForEach {$C \in S$ \atadd{in ascending order}}                                                                                              \labelline{state-transfer-for-begin}
        \State \Send{$\mUpdateRead$, $\iseqNumber$, $C$}{$\ireplicas(C)$}                                               \labelline{state-transfer-update-read}
        \State \WaitFor ($C \sqsubset \Ccurr$) $\lor$ (responses from any $Q \in \iquorums(C)$ with s.n. $\iseqNumber$) \labelline{state-transfer-wait}
      \EndFor                                                                                                           \labelline{state-transfer-for-end}
      \If {$\Ccurr \sqsubset \Cnext$}
        \State $\Ccurr \gets \Cnext$                                                                                    \labelline{state-transfer-ccurr-set}
        \State \URBBroadcastIn{$\mUpdateComplete$}{$\Cnext$}                                                            \labelline{state-transfer-broadcast}
        \State $\iactiveStateTransfer \gets \iFalse$
      \EndIf                                                                                                            \labelline{state-transfer-end}
    \EndHandler

    \algspace
    \UponRBDeliver{$\mNewHistory$, $h$, $\sigma$}{any sender}                                                           \labelline{new-history}
      \If {$\fVerifyHistory(h, \sigma) \land \ihistory \subset h$}                                                      \labelline{new-history-verify}
        \State $\ihistory \gets h$
        \State $\fUpdateFSKeys(\iheight(\fMaxElement(\ihistory)))$                                                      \labelline{new-history-update-fs-keys}
      \EndIf                                                                                                            \labelline{new-history-end}
    \EndHandler

    \algspace
    \UponReceive{$\mUpdateRead$, $\isn$, $C$}{replica $r'$}                                                             \labelline{upon-update-read}
      \State \WaitFor $C \sqsubset \fMaxElement(\ihistory)$ \Comment{only reply after $\fUpdateFSKeys$}                 \labelline{upon-update-read-wait}
      \State \Send{$\mUpdateReadResp$, $\icurrentValues$, $\isn$}{$r'$}                                                 \labelline{upon-update-read-send-resp}
    \EndHandler

    \algspace
    \UponReceive{$\mUpdateReadResp$, $\ivalues$, $\isn$}{replica $r'$}
      \If {$\fVerifyValues(\ivalues \setminus \icurrentValues)$} $\icurrentValues \gets \icurrentValues \cup \ivalues$ \EndIf
    \EndHandler

    \algspace
    \UponURBDeliverIn{$\mUpdateComplete$}{$C$}{quorum $Q \in \iquorums(C)$}                                             \labelline{upon-update-complete}
      \State \WaitFor $C \in \ihistory$                                                                                 \labelline{upon-update-complete-wait}
      \If {$\Cinst \sqsubset C$}                                                                                        \labelline{upon-update-complete-if}
        \If {$\Ccurr \sqsubset C$} $\Ccurr \gets C$ \EndIf                                                              \labelline{upon-update-complete-ccurr-set}
        \State $\Cinst \gets C$                                                                                         \labelline{upon-update-complete-cinst-set}
        \State \textbf{trigger} upcall $\fInstalledConfig(C)$                                                           \labelline{upon-update-complete-upcall}
        \If {$r \notin \ireplicas(C)$} halt \EndIf
      \EndIf                                                                                                            \labelline{upon-update-complete-end}
    \EndHandler
  \EndAlgorithmic
\end{algorithm*}

\atadd{The pseudocode for the replicas is presented in Algorithms~\lstref{dbla-replica-part-1} and~\lstref{dbla-replica-part-2}.}

Each replica $r$ maintains, locally, its \emph{current configuration} (denoted by $\Ccurr_r$)
and \emph{the last configuration installed by this replica} (denoted by $\Cinst_r$).
$\Cinst_r \sqsubseteq \Ccurr_r \sqsubseteq \Chighest_r$.
Intuitively, $\Ccurr_r = C$ means that replica $r$ knows that there is no need to transfer state
from configurations {\tlower} than $C$, either because $r$ already performed the state transfer from
those configurations, or because it knows that sufficiently many other replicas did.
$\Cinst_r = C$ means that the replica knows that sufficiently many replicas in $C$ have up-to-date states,
and that configuration $C$ is ready to serve user requests.

As we saw earlier, each client message is associated with some configuration $C$.
The replica only \atreplace{answers}{processes} the message when $C = \Cinst_r = \Ccurr_r = \Chighest_r$.
If $C \sqsubset \Chighest_r$, the replica simply ignores the message.
Due to the properties of reliable broadcast, the client will eventually learn about $\Chighest_r$ and
will repeat its request there (or in an even {\thigher} configuration).
If $\Cinst_r \sqsubset C$ and $\Chighest_r \sqsubseteq C$,
the replica waits until $C$ is installed before processing the message.
Finally, if $C$ is incomparable with $\Cinst_r$ or $\Chighest_r$, then,
\needsrev{since all candidate configurations are required to be comparable},
the message is sent by a Byzantine \atreplace{process}{client} and the replica should ignore it.

When a correct replica $r$ receives a {\mPropose} message \atadd{(\lstlineref{dbla-replica}{upon-propose})},
it adds the newly learned verifiable input values to $\icurVals_r$
and sends $\icurVals_r$ \needsrev{back} to the client with a forward-secure signature with timestamp $\iheight(C)$.
When a correct replica receives a {\mConfirm} message \atadd{(\lstlineref{dbla-replica}{upon-confirm})},
it simply signs the set of acknowledgments in it with a forward-secure signature with timestamp $\iheight(C)$
and sends the signature to the client.

A very important part of the replica's implementation is \emph{the state transfer protocol}.
The pseudocode for it is presented in Algorithm~\ref{alg:dbla-replica-part-2}.
\atremove{Note that we omit the subscript $r$ in the pseudocode because each process can only access its own variables directly.}
Let $\Cnext_r$ be the {\thighest} configuration in $\ihistory_r$ such that $r \in \ireplicas(\Cnext_r)$.
Whenever $\Ccurr_r \neq \Cnext_r$,
the replica tries to ``move'' to $\Cnext_r$ by reading the current state from all configurations between $\Ccurr_r$ and $\Cnext_r$ one by one in ascending order (\lstlineref{dbla-replica}{state-transfer-for-begin}).
In order to read the current state from configuration $C \sqsubset \Cnext_r$, replica $r$ sends
message \Message{$\mUpdateRead$, $\iseqNumber_r$, $C$} to all replicas in $\ireplicas(C)$.
In response, each replica $r_1 \in \ireplicas(C)$ sends $\icurVals_{r_1}$ to $r$ in an $\mUpdateReadResp$ message
(\lstlineref{dbla-replica}{upon-update-read-send-resp}).
However, $r_1$ replies only after its private key is updated to a timestamp larger than $\iheight(C)$
(\lstlineref{dbla-replica}{upon-update-read-wait}).
\atrev{We maintain the invariant (\lstlineref{dbla-replica}{new-history-update-fs-keys}) that for every correct replica $q$, the timestamp $\ist_{q}$ is always equal to $\iheight(\Chighest_q)$.}

If $r$ receives a quorum of replies from the replicas of $C$, there are two distinct cases:
\begin{itemize}
    \item 
        $C$ is still {\tactive} \atadd{at the moment when $r$ receives the last acknowledgment}.
        In this case, the quorum intersection property still holds for $C$, and replica $r$ can be sure that 
        (1) if some $\fPropose$ operation has \atremove{either} completed in configuration $C$ or reached the second stage \atadd{with some set of verifiable input values $\ivalues$}, then $vs \sqsubseteq \icurVals_r$;
        and (2) if some $\fPropose$ operation has not yet reached the second stage, it will not be able to complete in configuration $C$ (\atadd{it will have to retry in a {\thigher} configuration, }see the example in Figure~\ref{fig:slow-reader-demonstation}).
    \item
        $C$ is already {\tsuperseded} \atadd{by the time $r$ receives the last acknowledgment}.
        \atrev{This means that some configuration {\thigher} than $C$ is installed, and the state from configuration $C$ was already transferred to that {\thigher} configuration}.
        \atadd{We refer to Appendix~\ref{sec:dbla-correctness-proof} for more formal proofs.}
\end{itemize}

\atreplace{It may happen that configuration $C$ is already superseded and $r$ will not receive sufficiently many replies from the replicas of $C$.}
{If a configuration $C$ is {\tsuperseded} before $r$ receives enough replies, it may happen that $r$ will never be able to collect a quorum of replies from $C$.}
However, in this case, $r$ will eventually discover that some {\thigher} configuration is installed, and it will update $\Ccurr_r$ (\lstlineref{dbla-replica}{upon-update-complete-ccurr-set}).
\atadd{The waiting on line~\ref{lst:dbla-replica:state-transfer-wait} will terminate due to the first part of the condition ($C \sqsubset \Ccurr$).}

When a correct replica completes transferring the state to some configuration $C$,
it notifies other replicas about it by broadcasting message $\mUpdateComplete$ in configuration $C$
(\lstlineref{dbla-replica}{state-transfer-broadcast}).
A correct replica \emph{installs} a configuration $C$ if it receives such messages from a quorum of replicas in $C$
(\lstlineref{dbla-replica}{upon-update-complete}).
Because we want our protocol to satisfy the {\pInstallationLiveness} property
(if one correct replica installs a configuration, every forever-correct replica must eventually install this or a {\thigher} configuration),
the $\mUpdateComplete$ messages are distributed through the uniform reliable broadcast primitive that
we introduced in Section~\ref{subsec:bcast}.

\subsection{Time complexity}

\atrev{%
In our analysis we assume that the time complexity of the reliable broadcast primitive, which we use to disseminate verifiable histories, is constant.
With this assumption, it is easy to see that the worst-case time complexity of our {\DBLA} implementation is $O(m+k)$, where $m$ is the number of verifiable input values and $k$ is the size of the largest verifiable history.
Indeed, the time complexity is proportional to the number of calls to $\fRefine$.
There are only two reasons why a client may call $\fRefine$: either it learns about a new verifiable input value (\lstlineref{dbla-client}{upon-propose-resp-refine}), which may happen at most $m$ times or it learns about a new verifiable history (\lstlineref{dbla-client}{new-history-refine}), which may happen at most $k$ times.}

\atrev{%
As for the complexity of the state transfer protocol, it is linear in the size of the largest verifiable history $k$.
Because a replica advances the $\Ccurr$ variable at the end of state transfer (\lstlineref{dbla-replica}{state-transfer-ccurr-set}), each of the $k$ candidate configurations is accessed at most once by each replica, and in each configuration, our state transfer protocol makes a constant number of steps.}

\subsection{Implementing other {\dynamic} objects}

While we do not provide any general approach for building {\dynamic} objects, we expect that most asynchronous Byzantine fault-tolerant static algorithms can be adapted to the {\dynamic} case by applying the same set of techniques.
These techniques include our state transfer protocol
(relying on forward-secure signatures),
the use of an additional round-trip to prevent the ``slow reader'' attack,
and the structure of our cryptographic proofs ensuring that {\tentative} configurations cannot
create valid certificates for output values.
To illustrate this, in Appendix~\ref{sec:max-register},
we present the {\dynamic} version of {\MaxRegister}~\cite{max-register}.
We also discuss the {\dynamic} version of the {\AccessControl} abstraction in Section~\ref{sec:access-control}.

\section{Implementing reconfigurable objects} \label{sec:transform-main}

\begin{figure}
    \centering
    \pdftex[\linewidth]{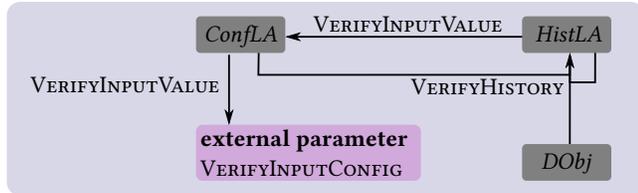}
 
    \caption{\atadd{The structure of dependencies in our implementation of a reconfigurable object. The arrow from object $A$ to object $B$ represents a dependency of object $A$ on object $B$. The label next to the arrow reflects the nature of this dependency.}}
 
  \label{fig:transform}
\end{figure}

While {\dynamic} objects are important building blocks, they are not particularly useful by themselves
because they require an external source of comparable verifiable histories.
%
In this section, we show how to combine several {\dynamic} objects to obtain a single
\emph{\reconfigurable} object.
Similar to {\dynamic} objects, the specification of a {\reconfigurable} object can be obtained as a combination of
the specification of a static object with the specification of an abstract {\reconfigurable} object from
Section~\ref{subsec:reconf-objects}.
In particular, compared to static objects, {\reconfigurable} objects have
one more operation -- $\fUpdateConfig(C, \sigma)$, must be parameterized
by a boolean function $\fVerifyInputConfig(C, \sigma)$, and must satisfy
{\pReconfigurationValidity}, {\pReconfigurationLiveness}, and {\pInstallationLiveness}.



We build a {\reconfigurable} object by combining three \emph{\dynamic} ones.
The first one is the {\dynamic} object that executes clients' operations (let us call it $\iDObj$).
\atrev{For example, in order to implement a {\reconfigurable} version of {\ByzantineLatticeAgreement}, one needs to take a {\dynamic} version of {\ByzantineLatticeAgreement} as $\iDObj$.
Similarly, in order to implement a {\reconfigurable} version of {\MaxRegister}~\cite{max-register}, one needs to take a {\dynamic} version of {\MaxRegister} as $\iDObj$ (see Appendix~\ref{sec:max-register}).}
The two remaining objects are used to build verifiable histories:
$\iConfLA$ is a {\DBLA} operating on the configuration lattice $\Conf$,
and $\iHistLA$ is a {\DBLA} operating on the powerset lattice $2^{\Conf}$.
The relationships between the three {\dynamic} objects are depicted in Figure~\ref{fig:transform}.

\begin{algorithm*}[!thb]
  \caption{{\Reconfigurable} object}
  \labelalg{transform}

  \BeginAlgorithmic
    \LineCommentx{Common code}
    \Parameters
      \State Lattice of configurations $\Conf$ and the initial configuration $\Cinit$
      \State Boolean function $\fVerifyInputConfig(C, \sigma)$
      \State Dynamic object $\iDObj$, which we want to make {\reconfigurable}
    \EndParameters

    \SharedObjects
      \State $\iDObj$     \Comment{the {\dynamic} object being transformed}
      \State $\iConfLA$   \Comment{{\DBLA} on lattice $\Conf$}
      \State $\iHistLA$   \Comment{{\DBLA} on lattice $2^{\Conf}$}
    \EndSharedObjects

    \algspace
    \LineCommentx{Code for client $p$}

    \State Data operations are performed directly on $\iDObj$.

    \algspace
    \Operation{$\fUpdateConfig$}{$C$, $\sigma$}
      \State let $\Tupple{D, \sigma_D} = \iConfLA.\fPropose(C, \sigma)$                                                 \labelline{update-config-conf-la-propose}
      \State let $\Tupple{h, \sigma_h} = \iHistLA.\fPropose(\{ D \}, \sigma_D)$                                         \labelline{update-config-hist-la-propose}
      \State $\iDObj.\fUpdateHistory(h, \sigma_h)$                                                                      \labelline{update-config-update-history-begin}
      \State $\iConfLA.\fUpdateHistory(h, \sigma_h)$
      \State $\iHistLA.\fUpdateHistory(h, \sigma_h)$                                                                    \labelline{update-config-update-history-end}
    \EndOperation

    \algspace
    \LineCommentx{Code for replica $r$}

    \Upon{receive upcall $\fInstalledConfig(C)$ from \textbf{all} $\iDObj$, $\iConfLA$, and $\iHistLA$}                 \labelline{installed-config-upcall}
      \State \textbf{trigger} upcall $\fInstalledConfig(C)$
    \EndHandler

    \algspace
    \LineCommentx{Parameters specification}
    \Function{$\iConfLA.\fVerifyInputValue$}{$v$, $\sigma$} = $\fVerifyInputConfig(v, \sigma)$ \EndFunction

    \Function{$\iHistLA.\fVerifyInputValue$}{$v$, $\sigma$}                                                             \labelline{histla-verify-input-value}
      \If {$v$ is not a set of 1 element} \Return $\iFalse$ \EndIf
      \State let $\{ C \} = v$
      \State \Return $\iConfLA.\fVerifyOutputValue(C, \sigma)$
    \EndFunction

    \LineCommentx{\atadd{All {\dynamic} objects are parameterized with the same {$\fVerifyHistory$} function.}}
    \Function{$\fVerifyHistory$}{$h$, $\sigma$} = $\iHistLA.\fVerifyOutputValue(h, \sigma)$
    \labelline{verify-history}
    \EndFunction
  \EndAlgorithmic
\end{algorithm*}


The pseudocode is presented in Algorithm~\lstref{transform}.
All data operations are performed directly on $\iDObj$.
To update a configuration, the client first submits its proposal to $\iConfLA$
and then submits the result as a singleton set to $\iHistLA$.
Due to the {\pBLAComparability} property, all verifiable output values
produced by $\iConfLA$ are comparable, and any combination of them would create
a well-formed history as defined in Section~\ref{subsec:dynamic-objects}.
Moreover, the verifiable output values of $\iHistLA$ are related by containment,
and, therefore, can be used as verifiable histories in {\dynamic} objects.
We use them to reconfigure all three {\dynamic} objects (lines~\ref{lst:transform:update-config-update-history-begin}--\ref{lst:transform:update-config-update-history-end}).

\myparagraph{Cryptographic keys.}
In Algorithm~\lstref{transform}, we use several {\dynamic} objects.
We assume that correct replicas have separate public/private key pairs for each dynamic object.
This prevents replay attacks across objects and allows each {\dynamic} object to
manage its keys separately.
We discuss how to avoid this assumption later in this section.

\subsection{Proof of correctness}

In the following two lemmas we show that we use the {\dynamic} objects ($\iConfLA$, $\iHistLA$, and $\iDObj$) correctly,
i.e., all requirements imposed on verifiable histories are satisfied.
\begin{lemma}
  All histories passed to the {\dynamic} objects by correct processes
  (\lstlinerangeref{transform}{update-config-update-history-begin}{update-config-update-history-end})
  are verifiable with $\fVerifyHistory$ (\lstlineref{transform}{verify-history}).
\end{lemma}
\begin{proof}
  Follows from the {\pBLAVerifiability} property of $\iHistLA$.
\end{proof}

\begin{lemma} \label{lem:transform-verifiable-histories}
  All histories verifiable with $\fVerifyHistory$ (\lstlineref{transform}{verify-history}) are
  (1)~well-formed (that is, consist of comparable configurations) and (2)~related by containment.
  Moreover, (3)~in any given infinite execution, there is only a finite number of histories verifiable with $\fVerifyHistory$.
\end{lemma}
\begin{proof}
  (1)~follows from the {\pBLAComparability} property of $\iConfLA$, the {\pBLAValidity} property of $\iHistLA$,
    and the definition of $\iHistLA.\fVerifyInputValue$ (\lstlineref{transform}{histla-verify-input-value}).

  (2)~follows directly from the {\pBLAComparability} property of $\iHistLA$.

  (3)~follows from the requirement of finite number of verifiable input configurations
    and the {\pBLAValidity} property of $\iConfLA$ and $\iHistLA$.
    \needsrev{%
    Only a finite number of configurations can be formed by $\iConfLA$ out of a finite number of verifiable input configurations,
    and only a finite number of histories can be formed by $\iHistLA$ out of the configurations produced by $\iConfLA$.}
\end{proof}

\begin{theorem}[Transformation safety] \label{the:transform-safety}
  Our implementation satisfies the {\pReconfigurationValidity} property of a {\reconfigurable} object.
  That is,
  (1) every installed configuration $C$ is a join of some set of verifiable input configurations;
  and (2) all installed configurations are comparable.
\end{theorem}
\begin{proof}
  (1) follows from the {\pBLAValidity} property of $\iConfLA$ and $\iHistLA$ and the {\pDynamicValidity} property of the underlying {\dynamic} objects.
  (2) follows directly from the {\pDynamicValidity} property of the underlying {\dynamic} objects.
\end{proof}


\begin{theorem}[Transformation liveness] \label{the:transform-liveness}
  Our implementation satisfies the liveness properties of a {\reconfigurable} object:
  {\pReconfigurationLiveness}
  and {\pInstallationLiveness}.
\end{theorem}
\begin{proof}
  {\pReconfigurationLiveness} follows from the {\pBLALiveness} property of $\iConfLA$ and $\iHistLA$
  and the {\pDynamicLiveness} property of the underlying {\dynamic} objects.
  {\pInstallationLiveness} follows from line~\ref{lst:transform:installed-config-upcall} of the implementation
  and the {\pInstallationLiveness} of the underlying {\dynamic} objects.
\end{proof}


\subsection{Discussion} \label{subsec:transform-discussion}

\myparagraph{Time complexity}

\atrev{%
By accessing $\iConfLA$ and then $\iHistLA$, we minimize the number of
configurations that should be accessed for a consistent configuration shift.
Indeed, due to the {\pBLAValidity} property of $\iConfLA$ and $\iHistLA$, when $k$
reconfiguration requests are executed concurrently, at most $k$ new verifiable histories will be created
and the total number of candidate configurations will not exceed $k+1$ (including the initial configuration).
As a result, only $O(k)$ configurations are accessed for state transfer, similar to~\cite{parsimonious} and~\cite{SKM17-reconf}. 
In contrast, in DynaStore~\cite{dynastore}, a client might have to access up to $\Omega(\min\{mk, 2^k\})$ configurations, where $m$ is the number of concurrent data operations.}

\atrev{%
Additionally, every reconfiguration request involves two invocations of $\text{\DBLA}.\fPropose$.
The worst-case latency of our {\DBLA} $\fPropose$ implementation is $O(k+m)$, where $m$ is the number of verifiable input values.
In this case, $m = k$.
%
Hence, the worst-case latency of a reconfiguration request is linear with respect to the number of verifiable input configurations, which is known to be optimal even for crash fault-tolerant systems~\cite{SKM17-reconf}.}%
\footnote{\atadd{The analogy for the number of verifiable input configurations in crash fault-tolerant systems is the number of reconfiguration requests. In the context of Byzantine fault-tolerant systems, we have to talk about the number of verifiable input configurations instead because a single Byzantine client can simulate an infinite sequence of requests.}}

\myparagraph{Bootstrapping}
The relationship between lattice agreement and reconfiguration has been studied before~\cite{smartmerge,rla}.
In particular, as shown in~\cite{smartmerge}, lattice agreement can be used to build comparable configurations.
We take a step further and use two separate instances of lattice agreement:
one to build comparable configurations ($\iConfLA$) and the other to build
histories out of them ($\iHistLA$).
These two LA objects can then be used to reconfigure a single {\dynamic} object ($\iDObj$).

However, this raises a question: how to reconfigure the lattice agreement objects themselves?
We found the answer in the idea that is sometimes referred to as ``bootstrapping''.
We use the lattice agreement objects to reconfigure \emph{themselves and at least one other object}.
This implies that the lattice agreement objects share the configurations with $\iDObj$.
The most natural implementation is that the code for all three dynamic objects \needsrev{($\iConfLA$, $\iHistLA$, and $\iDObj$)} will be executed by the same set of replicas.

Bootstrapping is a dangerous technique because, if applied badly, it can lead to infinite recursion.
However, we structured our solution in such a way that there is no recursion at all:
the client first makes normal requests to $\iConfLA$ and $\iHistLA$
and then uses the resulting history to reconfigure all {\dynamic} objects,
as if this history was obtained by the client from the outside of the system.
It is important to note that liveness of the call $\iHistLA.\fVerifyOutputValue(h, \sigma)$ is not
affected by reconfiguration: the function simply checks some digital signatures and is
guaranteed to always terminate given enough processing time.

\myparagraph{Shared parts}
All implementations of {\dynamic} objects presented in this paper have a similar structure.
For example, they all share the same state transfer implementation
(see Algorithm~\lstref{dbla-replica-part-2}).
However, we do not deny the possibility that other implementations of other {\dynamic}
objects might have very different implementations.
Therefore, in our transformation we use $\iDObj$ as a ``black box'' and do not make any assumptions about its implementation.
Moreover, for simplicity, we use the two {\DBLA} objects as ``black boxes'' as well.
In fact, $\iConfLA$ and $\iHistLA$ may have different implementations and the transformation will still work
as long as they satisfy the specification from Section~\ref{sec:definitions}.
However, this comes at a cost.

In particular, if implemented naively, a single {\reconfigurable} object will run several independent state transfer protocols, and a single correct replica will have several private/public key pairs (as mentioned earlier in this section).
But if, as in this paper, all {\dynamic} objects have similar implementations of their state transfer protocols, this can be done more efficiently by combining all state transfer protocols into one, which would need to transfer the states of all {\dynamic} objects and make sure that the superseded configurations are harmless.


\section{{\AccessControl}}\label{sec:access-control}

\needsrev{%
Our implementation of {\reconfigurable} objects relies on
the parameter function $\fVerifyInputConfig$.
Moreover, if we apply our transformation from Section~\ref{sec:transform-main}
to our implementation of {\DBLA} from Section~\ref{sec:dbla-main},
the resulting {\reconfigurable} object will rely on the parameter
function $\fVerifyInputValue$.
The implementation of these parameters is highly application-specific.
For example, in a storage system, it is reasonable to only allow requests
that modify some data if they are accompanied by a digital signature
of the owner of the data.
For the sake of completeness,
in this section, we present three generic implementations for these parameter functions.
We believe that each of these implementations is suitable for some applications.}

\needsrev{In order to do this,} we introduce the \emph{\AccessControl} object.
It exports one operation and one function:
\begin{itemize}
  \item Operation $\fRequestCert(v)$ returns a \emph{certificate} $\sigma$, which can be verified with $\fVerifyCert(v, \sigma)$,
    or the special value $\bot$, indicating that the permission was denied;

  \item Function $\fVerifyCert(v, \sigma)$ returns a boolean value.
\end{itemize}

The implementation of \emph{\AccessControl} must satisfy the following property:
\begin{itemize}
  \item \textit{\pCertificateVerifiability}:
    If $\fRequestCert(v)$ returned $\sigma$ to a correct process, then $\fVerifyCert(v, \sigma) = \iTrue$.
\end{itemize}

\needsrev{%
In Sections~\ref{subsec:ac-first}--\ref{subsec:ac-last}, we present three different implementations
of the {\dynamic} version of the {\AccessControl} object,
and in Section~\ref{subsec:combining-access-control-with-other-objects},
we show how to use it in order to implement the parameter functions $\fVerifyInputValue$ and $\fVerifyInputConfig$.}

\subsection{Trusted administrators} \label{subsec:ac-first}

A naive yet common approach to {\dynamic} systems is to have a \atadd{pre-configured} trusted administrator, who signs the reconfiguration requests.
However, if the administrator's private key is lost, the system might lose liveness, and if it is compromised, the system might lose even safety.
A more viable approach is to have $n$ administrators and to require $b+1$ of them to sign every certificate, for some $n$ and $b$ such that $0 \le b < n$.
In this case, the system will ``survive'' up to $b$ keys being compromised and up to $n-(b+1)$ keys being lost.

\atremove{An interesting problem, which we will not dive into here,
is to allow changing the set of administrators.
One possible solution would be to ``run a reconfiguration protocol'' among the administrators. 
Another approach is to include the set of administrators to the configuration lattice $\Conf$ of
the reconfigurable object itself and to use normal reconfiguration requests to change the
set of administrators.}

\subsection{Sanity-check approach} \label{subsec:ac-sanity-check}


\begin{algorithm*}
  \caption{\VoteBasedDynamicAccessControl}
  \labelalg{ac-short}

  \BeginAlgorithmic
    \LineCommentx{Code for client $p$}
    \Operation{$\fRequestCert$}{$v$}                                                                                    \labelline{request-cert}
      \State let $C = \fMaxElement(\ihistory)$                                                                          \labelline{request-cert-begin}
      \State $\iseqNumber \gets \iseqNumber + 1$   \Comment{used to match requests with responses}

      \vspace{0.5\baselineskip}
      \LineCommentx{Phase one: request}
      \State \Send{$\mRequest$, $v$, $\iseqNumber$, $C$}{$\ireplicas(C)$}
      \LineComment{``enough'' means $b+1$ (Section~\ref{subsec:ac-sanity-check}) or a quorum (Section~\ref{subsec:ac-quorum}).}
      \State \WaitFor ($\fMaxElement(\ihistory) \neq C$)
          $\lor$ (received enough Yes-votes with valid signatures)
          $\tlaStyleLor$ (received a quorum of votes in total) \labelline{request-cert-wait-one}
      \If {$\fMaxElement(\ihistory) \neq C$} restart the operation (\textbf{goto} line~\ref{lst:\algname:request-cert-begin}) \EndIf
      \If {received not enough valid Yes-votes} \Return $\bot$ \Comment{access denied} \EndIf
      \State let $\iacks_1 = \{ \text{Yes-votes received on line~\ref{lst:\algname:request-cert-wait-one}} \}$

      \vspace{0.5\baselineskip}
      \LineCommentx{Phase two: confirm}
      \State \Send{$\mConfirm$, $\iacks_1$, $\iseqNumber$, $C$}{$\ireplicas(C)$}
      \State \WaitFor ($\fMaxElement(\ihistory) \neq C$) $\lor$ (\atrev{a quorum of replies} with valid signatures) \labelline{request-cert-wait-two}
      \If {$\fMaxElement(\ihistory) \neq C$} restart the operation (\textbf{goto} line~\ref{lst:\algname:request-cert-begin}) \EndIf
      \State let $\iacks_2 = \{ \text{acknowledgments received on line~\ref{lst:\algname:request-cert-wait-two}} \}$

      \vspace{0.5\baselineskip}
      \LineCommentx{Return certificate}
      \State \Return $\Tupple{\ihistory, \Sigmahistory, \iacks_1, \iacks_2}$                                            \labelline{request-cert-end}
    \EndOperation

    \algspace
    \LineCommentx{Code for replica $r$}
    \UponReceive{$\mRequest$, $v$, $\isn$, $C$}{client $c$}                                                             \labelline{upon-request}
      \State \WaitFor $C = \Cinst \lor C \sqsubset \fMaxElement(\ihistory)$                                             \labelline{upon-request-wait}
      \If {$C = \fMaxElement(\ihistory)$}
        \If {$\fVoteYes(v)$} \Send{$\mYes$, $\fFSSign(\Tupple{\mYes, v, c}, \iheight(C))$, $\isn$}{$c$}
        \Else{} \Send{$\mNo$, $\isn$}{$c$} \EndIf
      \EndIf
    \EndHandler

    \algspace
    \UponReceive{$\mConfirm$, $\iacks$, $\isn$, $C$}{client $c$}                                                        \labelline{upon-confirm}
      \State \WaitFor $C \in \ihistory$
      \If {$C = \fMaxElement(\ihistory)$}
        \State let $\isignature = \fFSSign(\Tupple{\mConfirmResp, \iacks}, \iheight(C))$                                \labelline{upon-confirm-sign}
        \State \Send{$\mConfirmResp$, $\isignature$, $\isn$}{$c$}
      \EndIf
    \EndHandler
  \EndAlgorithmic
\end{algorithm*}

One of the simplest implementations of access control in a \emph{static} system is to require
at least $b+1$ replicas to sign each certificate,
where $b$ is the maximal possible number of Byzantine replicas,
sometimes called the \emph{resilience threshold}.
The correct replicas can perform some application-specific sanity checks before approving requests.

The key property of this approach is that it guarantees that
each valid certificate is signed by \emph{at least one correct replica}.
In many cases, this is sufficient to guarantee resilience both against the Sybil attacks~\cite{sybilattack}
and against attempts to flood the system with reconfiguration requests.
The correct replicas can check the identities of the new participants and refuse to sign excessively frequent requests.

In \emph{\dynamic} asynchronous systems, just providing $b+1$ signatures is not sufficient.
Despite the use of forward-secure signatures, in a superseded {\pivotal} configuration there might be
significantly more than $b$ Byzantine replicas with their private keys not removed (in fact, at least $2b$).
The straightforward way to implement this policy in a \emph{\dynamic} system is to add the \emph{confirming} phase,
as in our implementation of {\DynamicByzantineLatticeAgreement} (see Section~\ref{sec:dbla-main}),
after collecting $b+1$ signed approvals.
The confirming phase guarantees that, during the execution of the first phase, the configuration was {\tactive}.
The state transfer protocol should be the same as for {\DBLA} with the exception that no actual state is being transferred.
The only goal of the state \atadd{transfer} protocol in this case is \atremove{it} to make sure that the replicas update their private keys before a new configuration is installed.

This and the following approach can be generally described as ``vote-based'' access control policies.
The pseudocode for their {\dynamic} implementation is presented in Algorithm~\lstref{ac-short}.

\subsection{Quorum-based approach (``on-chain governance'')} \label{subsec:ac-quorum} \label{subsec:ac-last}

A more powerful strategy in a \emph{static} system is to require a \emph{quorum} of replicas to sign each certificate.
An important property of this implementation is that it can detect and prevent conflicting requests.
More formally, suppose that there are values $v_1$ and $v_2$, for which the following two properties should hold:
\begin{itemize}
  \item Both are acceptable: $\fRequestCert(v_i)$ should not return $\bot$ unless $\fRequestCert(v_j)$ was invoked in the same execution,
    where $j \neq i$.
  \item At most one may be accepted: if some process knows $\sigma_i$ such that $\fVerifyCert(v_i, \sigma_i)$
    then no process should know $\sigma_j$ such that $\fVerifyCert(v_j, \sigma_j)$.
\end{itemize}
Note that it is possible that neither $v_1$ nor $v_2$ is accepted by the {\AccessControl}
if the requests are made concurrently.
To guarantee that exactly one certificate is issued, we would need to implement consensus,
which is impossible in asynchronous model~\cite{flp}.
If a correct replica has signed a certificate for value $v_i$, it should store this fact
in persistent memory and refuse signing $v_j$ if requested.
Due to the quorum intersection property, this guarantees the ``at most one'' semantic in a static system.

This approach can be implemented in a \emph{\dynamic} system using the pseudocode from Algorithm~\lstref{ac-short}
and the state transfer protocol from our {\DBLA} implementation
(see Algorithm~\lstref{dbla-replica-part-2}).

Using the {\dynamic} version of this approach to certifying reconfiguration requests allows us to
capture the notion of what is sometimes called ``on-chain governance''.
The idea is that the participants of the system (in our case, the owners of the replicas) decide
which actions or updates to allow by the means of voting.
Every decision needs a quorum of signed votes to be considered valid
and no two conflicting decisions can be made.

\subsection{Combining {\AccessControl} with other objects} \label{subsec:combining-access-control-with-other-objects}

\begin{figure}
    \centering
    
    \begin{subfigure}{\linewidth}
        \centering
        \pdftex[\linewidth]{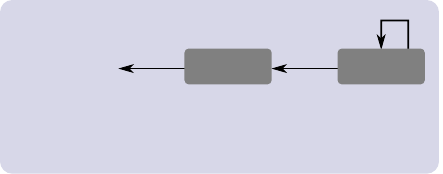}
        \caption{{\DynamicAccessControl} inside a {\reconfigurable} object.}
        \label{fig:access-control-inside}
    \end{subfigure}
    
    \begin{subfigure}{\linewidth}
        \centering
        \pdftex[\linewidth]{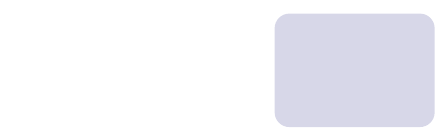}
        \caption{{\ReconfigurableAccessControl} in combination with another {\reconfigurable} object.}
        \label{fig:access-control-outside}
    \end{subfigure}
    
    \caption{%
        Two possible ways to integrate the {\AccessControl} abstraction with other types of objects.
        An arrow from an object $A$ to another object $B$ marked with $\fVIV$, (resp., $\fVIC$ or $\fVH$)
        indicates that $A.\fVerifyInputValue$ (resp., $A.\fVerifyInputConfig$ or $A.\fVerifyHistory$)
        is implemented using $B.\fVerifyOutputValue$ or $B.\fVerifyCert$.}
    \label{fig:access-control}
\end{figure}

There are at least two possible ways to combine the {\AccessControl} abstraction with
a {\reconfigurable} object in a practical system.

The simplest, and, perhaps, the most practical approach is to embed two instances of
{\DynamicAccessControl} directly into the structure of a reconfigurable object,
as shown in Figure~\ref{fig:access-control-inside}.
In this case, the replicas that execute the code for the {\AccessControl} are the same
replicas as the replicas that execute the code of other {\dynamic} objects in the implementation
of this {\reconfigurable} object.

Alternatively, one can apply the transformation from Section~\ref{sec:transform-main} to the {\dynamic}
{\AccessControl} implementation described in this section to obtain a \emph{\reconfigurable}
version of the {\AccessControl} abstraction.
It then can be combined with any other {\reconfigurable} object in a structure depicted in Figure~\ref{fig:access-control-outside}.
In this case, the replicas of $\iConfRAC$ produce verifiable input configurations for themselves and for two other objects.

\section{Related work}\label{sec:related-work}

Dynamic replicated systems with \emph{passive
  reconfiguration}~\cite{BaldoniBKR09,AttiyaCEKW19,KW19} do not explicitly
regulate arrivals and departures of replicas.
Their consistency properties are ensured under strong
assumptions on the churn rate.
Except for the recent work~\cite{KW19}, churn-tolerant storage systems do not tolerate Byzantine failures.
In contrast, \textit{active reconfiguration} allows the clients to explicitly propose
configuration updates, e.g., sets of new replica arrivals and departures. 

Early proposals of (actively) reconfigurable storage systems
tolerating process crashes, such as RAMBO~\cite{rambo} and
reconfigurable Paxos~\cite{paxos-reconfigure},
used consensus (and, thus, assumed certain level of synchrony) to ensure that the clients agree on the evolution of configurations.
DynaStore~\cite{dynastore} was the first \emph{asynchronous} reconfigurable storage:
clients propose incremental additions or removals
to the system configuration.
As the proposals commute, the processes can resolve their disagreements without
involving consensus.

The \textit{parsimonious speculative snapshot}
task~\cite{parsimonious} resolves conflicts between
concurrent configuration updates in a storage system using instances
of commit-adopt~\cite{Gaf98}.
The worst-case time complexity, in the number of message delays, of
reconfiguration was later reduced from $O(nm)$~\cite{parsimonious} to
$O(n+m)$~\cite{SKM17-reconf}, where $n$ is the number of concurrently
proposed configuration updates and $m$ is the number of concurrent data operations.
\atrev{This coincides with the time complexity of our solution.}

SmartMerge~\cite{smartmerge} made an important step forward by
treating reconfiguration as an instance of abstract \emph{lattice
agreement}~\cite{gla}.
However, the algorithm assumes an external (reliable) lattice agreement service
which makes the system not fully reconfigurable.

\pkadd{FreeStore~\cite{freestore} describes an algorithm for reconfigurable storage that can be seen as a composition of a reconfiguration protocol and a read-write protocol. 
Reconfiguration is based on the \emph{view generator} abstraction\atreplace{ that}{, which} encapsulates the form of agreement required to reconcile concurrently proposed reconfiguration requests (essentially, very \atreplace{close}{similar} to lattice agreement).
The use of view generators helps \atadd{in} optimizing \atremove{communication} latency, similar to our use of dynamic lattice agreement objects.
\atremove{\atreplace{Configuration}{Reconfiguration} requests in FreeStore are restricted to individual \emph{join} and \emph{leave} operations.}}

The recently proposed \atremove{the} \emph{reconfigurable lattice-agreement}
abstraction~\cite{rla} enables  reconfigurable versions of a large class of
objects and constructions, including state-based CRDTs~\cite{crdt}, atomic-snapshot, max-register,
conflict detector and commit-adopt.
Configurations are treated here in an abstract way, as elements of a configuration lattice, encapsulating replica sets and quorum assumptions. 
We believe that the reconfiguration service we introduced in this
paper can be used to derive Byzantine fault-tolerant reconfigurable
implementations of objects in the class.

Byzantine quorum systems~\cite{ByzantineQuorumSystems} introduce
abstractions for ensuring availability and consistency of shared data in asynchronous systems
with Byzantine faults. In particular, a \emph{dissemination} quorum system
ensures that every two quorums have a correct process in common
and that at least one quorum only contains correct processes.
%

Dynamic Byzantine quorum systems~\cite{alvisi2000dynamic} appear to be the
first attempt to implement a form of active reconfiguration in a
Byzantine fault-tolerant data service running on a \emph{static} set of
replicas, where clients can raise or lower the resilience threshold.
Dynamic Byzantine storage~\cite{martin2004framework} allows a trusted
\emph{administrator} to issue ordered reconfiguration
calls that might also change the set of replicas.
The administrator is also responsible for generating new private keys for
the replicas in each new configuration to anticipate the {\attackName} attack~\cite{tutorial2010}.
%
In this paper, we propose an implementation of a Byzantine
fault-tolerant reconfiguration service that does not rely on this
assumption.

%
\atrev{Forward-secure signature schemes~\cite{bellare1999forward,boyen2006forward,canetti2007forward,drijvers2019pixel,malkin2002efficient}
were originally designed to mitigate the consequences of key exposure:
if the private key of an agent is compromised,
signatures made prior to the exposure (i.e., with smaller timestamps) can still be trusted.
In this paper, a novel application of forward-secure digital signatures is proposed: timestamps are associated with configurations. Before a new configuration is installed, the protocol ensures that sufficiently many correct processes update their private keys in prior configurations.
This approach prevents the ``I still work here'' and ``slow reader'' attacks.
Unlike previously proposed solutions~\cite{martin2004framework}, it does not rely on a global agreement on the configuration sequence or a trusted administrator.}



\section{Discussion} \label{sec:discussion}

\subsection{Possible optimizations} \label{subsec:dbla-optimizations}

\atrev{In this paper, our goal was to provide the minimal implementation for the minimal set of abstractions to demonstrate the ideas and the general techniques for defining and building reconfigurable services in the harsh world of asynchrony and Byzantine failures.
Therefore, our implementations leave plenty of space for optimizations.
Here we would like to mention a few possible directions.
Most of them are dedicated to reducing the communication cost of the protocol.}

First, the proofs in our protocol include the full local history of a process.
Moreover, this history comes with its own proof, which also usually contains a history, and so on.
If implemented naively, the size of one proof in bytes will be at least quadratic with respect to the number of distinct candidate configurations, which is \atreplace{completely unnecessary}{not necessary}.
The first observation is that these histories will be related by containment.
So, in fact, they can be compressed just to the size of the largest one, which is linear.
But we can go further and say that, in fact, in a practical implementation, the processes almost never should actually send full histories to each other because every process maintains its local history and all histories with proofs are already disseminated via the reliable broadcast primitive.
When one process wants to send some history to some other process, it can just send a cryptographic hash of this history.
The other process can check if it already has this history and, if not, ask the sender to only send the missing parts, instead of the whole history.

Second, a naive implementation of our {\DBLA} protocol would send ever-growing sets of verifiable input values, which is\atreplace{, just as with histories, completely unnecessary}{ also not necessary}.
The processes should just limit themselves to sending diffs between what they know and what they think the recipient knows.

Third, almost every proof in our systems contains signatures from a quorum of replicas.
This adds another linear factor to the communication cost.
However, it can be significantly reduced by the use of forward-secure \emph{multi-signatures}, such as Pixel~\cite{drijvers2019pixel}, which was designed for similar \atreplace{purposes}{applications}.

Finally, we use a suboptimal implementation of lattice agreement as the foundation for our {\DBLA} protocol.
Perhaps, we could benefit from adapting a more efficient crash fault-tolerant asynchronous solution~\cite{garg-la}.

\myparagraph{Open questions.}
\oldconfirm{We would like to mention two relevant directions for further research.}

First, with regard to active reconfiguration, it would be interesting to devise algorithms that efficiently
adapt to ``small'' configuration changes, while still supporting
the option of completely changing the set of replicas in a single reconfiguration request.
In this paper, we allow
\atreplace{the sets of replicas of proposed configurations to be completely disjoint}{each reconfiguration request to completely change the set of replicas},
which \atrev{leads to} an expensive quorum-to-quorum communication pattern.
This seems unnecessary for reconfiguration requests involving only slight changes to the set of replicas.

Second, with regard to Byzantine faults, it would be interesting to consider models with a ``weaker'' adversary.
In this paper, we assumed a very strong model of the adversary: no assumptions are made about correctness
of replicas in superseded configurations.
This ``pessimistic'' approach leads to more complicated and expensive protocols.

\section*{Acknowledgments}
This work was supported \atadd{in part} by TrustShare Innovation Chair. We would also like to thank the anonymous reviewers from the DISC program committee and the Distributed Computing Journal board for their constructive comments and suggestions.

\bibliographystyle{ACM-Reference-Format}
\bibliography{main}

\clearpage
\appendix

\section{Proof of correctness of {\DBLA}} \label{sec:dbla-correctness-proof}

\subsection{Safety}

Recall that a configuration is called \emph{candidate} iff it appears in some verifiable history.
The following lemma gathers some obvious yet very useful statements about candidate configurations.
\begin{lemma}[Candidate configurations] \label{lem:candidate-configurations} \leavevmode
  \begin{enumerate}
    \item \label{itm:cand-fin}      There is a finite number of candidate configurations.
    \item \label{itm:cand-comp}     All candidate configurations are comparable with ``$\sqsubseteq$''. 
  \end{enumerate}
\end{lemma}
\begin{proof}
  %
  The total number of verifiable histories is required to be finite, and each history is finite, hence (\ref{itm:cand-fin}).
  All verifiable histories are required to be related by containment and all configurations within one history are
  required to be comparable, hence (\ref{itm:cand-comp}).
\end{proof}

Recall that a configuration is called \emph{\pivotal} if it is the last configuration in some verifiable history.
Non-{\pivotal} candidate configurations are called \emph{\tentative}.
Intuitively, the next lemma states that in the rest of the proofs we can almost always
consider only {\pivotal} configurations.
{\Tentative} configurations are both harmless and useless.
\begin{lemma}[{\Tentative} configurations] \label{lem:tentative-configurations} \leavevmode
  \begin{enumerate}
    \item No correct client will ever make a request to a {\tentative} configuration.
    \item {\Tentative} configurations cannot be installed.
    \item A correct process will never invoke $\fFSVerify$ with timestamp $\iheight(C)$ for any {\tentative} configuration $C$.
    \item A correct replica will never broadcast any message via the uniform reliable broadcast primitive in a {\tentative} configuration.
  \end{enumerate}
\end{lemma}
\begin{proof}
  Follows directly from the algorithm.
  Both clients and replicas only operate on configurations that were obtained by invoking the function $\fMaxElement(h)$
  on some verifiable configuration.
\end{proof}

The next lemma states that correct processes cannot ``miss'' any \emph{\pivotal} configurations in their local histories.
This is crucial for the correctness of the state transfer protocol.
\begin{lemma} \label{lem:pivotal-configurations}
  If $C \sqsubseteq \fMaxElement(h)$, where $C$ is a {\pivotal} configuration
  and $h$ is the local history of a correct process,
  then $C \in h$.
\end{lemma}
\begin{proof}
  Follows directly from the definition of a {\pivotal} configuration and
  the requirement that all verifiable histories are related by containment (see Section~\ref{subsec:dynamic-objects}).
\end{proof}

Recall that a configuration is called \emph{superseded} iff some {\thigher} configuration is installed (see Section~\ref{subsec:reconf-objects}).
A configuration is \emph{installed} iff some \emph{correct} replica has triggered the $\fInstalledConfig$ upcall
(\lstlineref{dbla-replica}{upon-update-complete-upcall}).
For this, the correct replica must receive a quorum of $\mUpdateComplete$ messages via the uniform reliable broadcast primitive
(\lstlineref{dbla-replica}{upon-update-complete}).

\begin{theorem}[\pDynamicValidity] \label{the:dynamic-validity}
  Our implementation of {\DBLA} satisfies {\pDynamicValidity}.
  I.e., only a candidate configuration can be installed.
\end{theorem}
\begin{proof}
  Follows directly from the implementation.
  A correct replica will not install a configuration until it is in the replica's local history
  (\lstlineref{dbla-replica}{upon-update-complete-wait}).
\end{proof}


\needsrev{%
In our algorithm, it is possible for a configuration to be installed \emph{after} it was superseded.
Imagine that a quorum of replicas broadcast $\mUpdateComplete$ messages in some configuration $C$ which is not yet installed.
After that, before any replica delivers those messages, a {\thigher} configuration is installed, making $C$ superseded.
It is possible that some correct replica $r \in \ireplicas(C)$ that does not yet know that a {\thigher} configuration is installed,
will deliver the broadcast messages and trigger the upcall $\fInstalledConfig(C)$ (\lstlineref{dbla-replica}{upon-update-complete-upcall}).}

\needsrev{%
Let us call the configurations that were installed while being {\tactive} (i.e., not superseded) ``\emph{properly installed}''.
We will use this definition to prove next few lemmas.}

\begin{lemma}
  The {\tlowest} properly installed configuration {\thigher} than configuration $C$ is
  the first installed configuration {\thigher} than $C$ \needsrev{in the real-time order}.  
\end{lemma}
\begin{proof}
  Let $N$ be the {\tlowest} properly installed configuration {\thigher} than $C$.
  If some configuration {\thigher} than $N$ were installed earlier,
  then $N$ would not be properly installed \needsrev{(by the definition of a properly installed configuration)}.
  If some configuration between $C$ and $N$ were installed earlier,
  then $N$ would not be the {\tlowest}.
\end{proof}

The following lemma stipulates that our state transfer protocol makes the superseded
{\pivotal} configurations ``harmless'' by leveraging a forward-secure signature scheme.
\begin{lemma}[Key update] \label{lem:key-update}
  If a {\pivotal} configuration $C$ is superseded,
  then no quorum of replicas in that configuration is capable of signing messages with timestamp $\iheight(C)$,
  i.e., $\nexists Q \in \iquorums(C) \textST \forall r \in Q: \ist_r \le \iheight(C)$.
\end{lemma}
\begin{proof}
  Let $N$ be the {\tlowest} properly installed configuration {\thigher} than $C$.
  Let us consider the moment when $N$ was installed.
  By the algorithm, all correct replicas in some quorum $Q_N \in \iquorums(N)$
  had to broadcast $\mUpdateComplete$ messages before $N$ was installed
  (\lstlineref{dbla-replica}{upon-update-complete}).
  Since $N$ was not yet superseded at that moment, there was at least one correct replica $r_N \in Q_N$.

  By Lemma~\ref{lem:pivotal-configurations}, $C$ was in $r_N$'s
  local history whenever it performed state transfer to any configuration {\thigher} than $C$.
  By the protocol, a correct replica only advances its $\Ccurr$ variable after executing the state transfer protocol
  (\lstlineref{dbla-replica}{state-transfer-ccurr-set})
  or right before installing a configuration
  (\lstlineref{dbla-replica}{upon-update-complete-ccurr-set}).
  Since no configurations between $C$ and $N$ were yet installed,
  $r_N$ had to pass through $C$ in its state transfer protocol and to receive
  $\mUpdateReadResp$ messages from some quorum $Q_C \in \iquorums(C)$
  (\lstlineref{dbla-replica}{state-transfer-wait}).

  Recall that correct replicas update their private keys whenever they learn about a {\thigher} configuration
  (\lstlineref{dbla-replica}{new-history-update-fs-keys})
  and that they will only reply to message $\Tupple{\mUpdateRead, \isn, C}$ once $C$ is \emph{not} the {\thighest}
  configuration in their local histories
  (\lstlineref{dbla-replica}{upon-update-read-wait}).
  This means that all correct replicas in $Q_C$ actually had to update their private keys before $N$
  was installed, and, hence, before $C$ was superseded.
  By the quorum intersection property, this means that in each quorum in $C$ at least one replica
  updated its private key to a timestamp higher than $\iheight(C)$ and will not
  be capable of signing messages with timestamp $\iheight(C)$ even if it becomes Byzantine.
\end{proof}

Note that in a {\tentative} configuration there might be arbitrarily many Byzantine replicas that have not updated their private keys.
This is inevitable in asynchronous system: forcing the replicas in {\tentative} configurations to update their private keys would
require solving consensus.
This does not affect correct processes because, as shown in Lemma~\ref{lem:tentative-configurations},
{\tentative} configurations are harmless.
However, it is important to remember this when designing new {\dynamic} protocols.

The following lemma implies that the state is correctly transferred between configurations.
\begin{lemma}[State transfer correctness] \label{lem:state-transfer-correctness} \leavevmode \\
  If $\sigma = \Tupple{\ivalues, h, \sigma_h, \iproposeAcks, \iconfirmAcks}$ is a valid proof for $v$,
  then for each {\tactive} installed configuration $D$ such that $\fMaxElement(h) \sqsubset D$,
  there is a quorum $Q_D \in \iquorums(D)$ such that for each correct replica $r \in Q_D$:
  $\ivalues \subseteq \icurrentValues_r$.
\end{lemma}
\begin{proof}
  Let $C = \fMaxElement(h)$.
  We proceed by induction on the sequence of all properly installed configurations {\thigher} than $C$.
  Let us denote this sequence by $\Cnice$.
  \needsrev{By the definition of a properly installed configuration, these are precisely the configurations that we
  consider in the statement of the lemma.} 

  Let $N$ be the {\tlowest} configuration in $\Cnice$.
  Let $Q_C \in \iquorums(C)$ be a quorum of replicas whose signatures are in $\iproposeAcks$.
  \needsrev{%
  Consider the moment of installation of $N$.
  There must be a quorum $Q_N \in \iquorums(N)$ in which all correct replicas
  broadcast $\Message{\mUpdateComplete, N}$ before the moment of installation.}
  For each correct replica $r_N \in Q_N$,
  $r_N$ passed with its state transfer protocol through configuration $C$
  and received $\mUpdateReadResp$ messages from some quorum of replicas in $C$.
  Note that at that moment configuration $C$ was not yet superseded.
  By the quorum intersection property,
  there is at least one correct replica $r_C \in Q_C$ that sent an $\mUpdateReadResp$ message to $r_N$
  (\lstlineref{dbla-replica}{upon-update-read-send-resp}).
  \needsrev{Since $r_C$ will send the $\mUpdateReadResp$ message only after updating its private keys}
  (\lstlineref{dbla-replica}{upon-update-read-wait}),
  it had to sign $\Tupple{\mProposeResp, \ivalues}$
  (\lstlineref{dbla-replica}{upon-propose-sign})
  before sending reply to $r_N$,
  which means that the $\mUpdateReadResp$ message from $r_C$ to $r_N$
  must have contained a set of values that includes all values from $\ivalues$.
  This proves the base case of the induction.

  Let us consider any configuration $D \in \Cnice$ such that $N \sqsubset D$.
  Let $M$ be the {\thighest} configuration in $\Cnice$ such that $N \sqsubseteq M \sqsubset D$
  (in other words, the closest to $D$ in $\Cnice$).
  Assume that the statement holds for $M$,
  i.e., while $M$ was {\tactive}, there were a quorum $Q_M \in \iquorums(M)$
  such that for each correct replica $r_M \in Q_M$: $\ivalues \subseteq \icurrentValues_{r_M}$.
  Similarly to the base case,
  let us consider a quorum $Q_D \in \iquorums(D)$ such that every correct replica in $Q_D$
  reliably broadcast $\Message{\mUpdateComplete, D}$ before $D$ was installed.
  For each correct replica $r_D \in Q_D$,
  by the quorum intersection property,
  there is at least one correct replica in $Q_M$ that sent an $\mUpdateReadResp$ message to $r_D$.
  This replica attached its $\icurrentValues$ to the message, which contained $\ivalues$.
  This proves the induction step and completes the proof.
\end{proof}

The next lemma states that if two output values were produced in the same configuration, they are comparable.
In a static system it could be proven by simply referring to the quorum intersection property.
In a {\dynamic} Byzantine system, however, to use the quorum intersection, we need to prove that the
configuration was {\tactive} during the whole period when the clients were exchanging data with the replicas.
In other words, we need to prove that the ``slow reader'' attack is impossible.
Luckily, we have the second stage of our algorithm designed for this sole purpose.
\begin{lemma}[{\pBLAComparability} in one configuration] \label{lem:comparable-in-one-config} \leavevmode \\
  If $\sigma_1 = \Tupple{\ivalues_1, h_1, \sigma_{h1}, \iproposeAcks_1, \iconfirmAcks_1}$ is a valid proof for output value $v_1$, \\
  and $\sigma_2 = \Tupple{\ivalues_2, h_2, \sigma_{h2}, \iproposeAcks_2, \iconfirmAcks_2}$ is a valid proof for output value $v_2$, \\
  and $\fMaxElement(h_1) = \fMaxElement(h_2)$,
  then $v_1$ and $v_2$ are comparable.
\end{lemma}
\begin{proof}
  Let $C = \fMaxElement(h_1) = \fMaxElement(h_2)$.
  By definition, the fact that $\sigma$ is a valid proof for $v$ implies that $\fVerifyOutputValue(v, \sigma) = \iTrue$
  (\lstlineref{dbla-client}{verify-output-value}).
  By the implementation, $h_1$ and $h_2$ are verifiable histories
  (\lstlineref{dbla-client}{verify-output-value-verify-history}).
  Therefore, $C$ is a {\pivotal} configuration.

  The set $\iconfirmAcks_1$ contains signatures from a quorum of replicas of configuration $C$, with timestamp $\iheight(C)$.
  Each of these signatures had to be produced after each of the signatures in $\iproposeAcks_1$
  because they sign the message $\Tupple{\mConfirmResp, \iproposeAcks_1}$
  (\lstlineref{dbla-replica}{upon-confirm-sign}).
  Combining this with the statement of Lemma~\ref{lem:key-update} (Key Update),
  it follows that at the moment when the last signature in the set $\iproposeAcks_1$ was created,
  the configuration $C$ was {\tactive} (otherwise it would be impossible to gather $\iconfirmAcks_1$).
  We can apply the same argument to the sets $\iproposeAcks_2$ and $\iconfirmAcks_2$.

  It follows that there are quorums $Q_1, Q_2 \in \iquorums(C)$
  and a moment in time $t$ such that:
  (1) $C$ is not superseded at time $t$,
  (2) all \emph{correct} replicas in $Q_1$ signed message $\Tupple{\mProposeResp, \ivalues_1}$ before $t$,
  and (3) all \emph{correct} replica in $Q_2$ signed message $\Tupple{\mProposeResp, \ivalues_2}$ before $t$.
  Since $C$ is not superseded at time $t$, there must be a correct replica in $Q_1 \cap Q_2$
  (due to {\QuorumIntersection}), which signed both $\Tupple{\mProposeResp, \ivalues_1}$
  and $\Tupple{\mProposeResp, \ivalues_2}$
  (\lstlineref{dbla-replica}{upon-propose-sign}).
  Since correct replicas only sign $\mProposeResp$ messages with comparable sets of values\footnote{
    Indeed, set $\icurrentValues$ at each correct replica can only grow,
    and the replicas only sign messages with the same set of verifiable input values as $\icurrentValues$
    (see \lstlinerangeref{dbla-replica}{upon-propose-cur-vals}{upon-propose-sign}).
  },
  $\ivalues_1$ and $\ivalues_2$ are comparable,
  i.e., either $\ivalues_1 \subseteq \ivalues_2$ or $\ivalues_2 \subset \ivalues_1$.
  Hence, $v_1 = \fJoinAll(\ivalues_1)$ and $v_2 = \fJoinAll(\ivalues_2)$ are comparable.
\end{proof}

Finally, let us combine the two previous lemmas to prove the {\pBLAComparability} property of our {\DBLA} implementation.
\begin{theorem}[{\pBLAComparability}] \label{the:comparability}
  Our implementation of {\DBLA} satisfies the {\pBLAComparability} property.
  That is, all verifiable output values are comparable.
\end{theorem}
\begin{proof}
  Let $\sigma_1 = \Tupple{\ivalues_1, h_1, \sigma_{h1}, \iproposeAcks_1, \iconfirmAcks_1}$ be a valid proof for output value $v_1$,
  and $\sigma_2 = \Tupple{\ivalues_2, h_2, \sigma_{h2}, \iproposeAcks_2, \iconfirmAcks_2}$ be a valid proof for output value $v_2$.
  Also, let $C_1 = \fMaxElement(h_1)$ and $C_2 = \fMaxElement(h_2)$.
  Since $h_1$ and $h_2$ are verifiable histories
  (\lstlineref{dbla-client}{verify-output-value-verify-history}),
  both $C_1$ and $C_2$ are {\pivotal} by definition.

  If $C_1 = C_2$, $v_1$ and $v_2$ are comparable by Lemma~\ref{lem:comparable-in-one-config}.

  Consider the case when $C_1 \neq C_2$.
  Without loss of generality, assume that $C_1 \sqsubset C_2$.
  Let $Q_1 \in \iquorums(C_2)$ be a quorum of replicas whose signatures are in $\iproposeAcks_2$.
  Let $t$ be the moment when first correct replica signed $\Message{\mProposeResp, \ivalues_2}$.
  Correct replicas only start processing user requests in a configuration when this configuration is installed
  (\lstlineref{dbla-replica}{upon-propose-wait}).
  Therefore, by Lemma~\ref{lem:state-transfer-correctness},
  at time $t$ there was a quorum of replicas $Q_2 \in \iquorums(C_2)$
  such that for every correct replica in $Q_2$: $\ivalues_1 \subseteq \icurrentValues$.
  By the quorum intersection property, there must be at least one correct replica in $Q_1 \cap Q_2$.
  Hence, $\ivalues_1 \subseteq \ivalues_2$ and $\fJoinAll(\ivalues_1) \sqsubseteq \fJoinAll(\ivalues_2)$.
\end{proof}
%

\begin{theorem}[{\DBLA} safety] \label{the:dbla-safety}
  Our implementation satisfies the safety properties of {\DBLA}:
  {\pBLAValidity},
  {\pBLAVerifiability},
  {\pBLAInclusion},
  {\pBLAComparability},
  and {\pDynamicValidity}.
\end{theorem}
\begin{proof} \leavevmode
  \begin{itemize}
    \item {\pBLAValidity}
      follows directly from the implementation: a correct client collects verifiable input values and joins them
      before returning from $\fPropose$ (\lstlineref{dbla-client}{propose-return});
    \item {\pBLAVerifiability} follows directly from how correct replicas form and check the proofs for output values
      (lines~\ref{lst:dbla-client:propose-make-proof}
      and~\ref{lst:dbla-client:verify-output-value-unpack}--\ref{lst:dbla-client:verify-output-value-end});
    \item {\pBLAInclusion} follows from the fact that the set $\icurrentValues$ of a correct client only grows
      (\lstlineref{dbla-client}{refine-current-values});
    \item {\pBLAComparability} follows from Theorem~\ref{the:comparability};
    \item Finally, {\pDynamicValidity} follows from Theorem~\ref{the:dynamic-validity}.
  \end{itemize}
\end{proof}

\subsection{Liveness}

\begin{lemma}[History Convergence] \label{lem:history-convergence}
  Local histories of all correct processes will eventually become identical.
\end{lemma}
\begin{proof}
  Let $p$ and $q$ be any two forever-correct processes\footnote{
    If either $p$ or $q$ eventually halts or becomes Byzantine,
    their local histories are not required to converge.
  }.
  Suppose, for contradiction, that the local histories of $p$ and $q$
  have diverged at some point and will never converge again.
  Recall that correct processes only adopt verifiable histories,
  and that we require the total number of verifiable histories to be finite.
  Therefore, there is some history $h_p$,
  which is the the largest history ever adopted by $p$,
  and some history $h_q$
  which is the the largest history ever adopted by $q$.
  Since all verifiable histories are required to be related by containment,
  and we assume that $h_p \neq h_q$, one of them must be a subset of the other.
  Without loss of generality, suppose that $h_p \subset h_q$.
  Since $q$ had to deliver $h_q$ through reliable broadcast
  (unless $h_q$ is the initial history) and $q$ remains correct
  forever, $p$ will eventually deliver $h_q$ as well, and
  will adopt it.
  \needsrev{Hence, $h_p$ is not the largest history ever adopted by $p$.}
  A contradiction.
\end{proof}

Next, we introduce an important definition, which we will use throughout the rest of the proofs.
\begin{definition}[Maximal installed configuration] \label{def:cmax}
  In a given infinite execution,
  a \emph{maximal installed configuration} is a configuration that eventually becomes installed
  and never becomes superseded.
\end{definition}

\begin{lemma}[$\Cmax$ existence] \label{lem:cmax-existence}
  In any infinite execution there is a unique maximal installed configuration.
\end{lemma}
\begin{proof}
  By Lemma~\ref{lem:candidate-configurations} (Candidate configurations)
  and Theorem~\ref{the:dynamic-validity} (\pDynamicValidity),
  the total number of installed configurations is finite and they are comparable.
  Hence, we can choose a unique maximum among them, which is never superseded by definition.
\end{proof}

Let us denote the (unique) maximal installed configuration by $\Cmax$.
\begin{lemma}[$\Cmax$ installation] \label{lem:cmax-installation}
  The maximal installed configuration will eventually be installed by all correct replicas.
\end{lemma}
\begin{proof}
  Since $\Cmax$ is installed, by definition, at some point some correct replica has
  triggered upcall $\fInstalledConfig(\Cmax)$
  (\lstlineref{dbla-replica}{upon-update-complete-upcall}).
  This, in turn, means that this replica delivered a quorum of $\mUpdateComplete$ messages
  via the \emph{uniform reliable broadcast} in $\Cmax$ when it was correct.
  Therefore, even if this replica later becomes Byzantine,
  by definition of the uniform reliable broadcast,
  either $\Cmax$ will become superseded (which is impossible),
  or every correct replica will eventually deliver the same set of $\mUpdateComplete$ messages
  and install $\Cmax$.
\end{proof}

\begin{lemma}[State transfer progress] \label{lem:state-transfer-progress}
  State transfer
  (\lstlinerangeref{dbla-replica}{state-transfer-begin}{state-transfer-end})
  executed by a forever-correct replica always terminates.
\end{lemma}
\begin{proof}
  Let $r$ be a correct replica executing state transfer.
  By Lemma~\ref{lem:candidate-configurations}, the total number of candidate configurations is finite.
  Therefore, it is enough to prove that there is no such configuration that $r$
  will wait for replies from a quorum of that configuration indefinitely
  (\lstlineref{dbla-replica}{state-transfer-wait}).
  %
  Suppose, for contradiction, that there is such configuration $C$.

  If $C \sqsubset \Cmax$,
  then, by Lemma~\ref{lem:cmax-installation}, $r$ will eventually install $\Cmax$,
  and $\Ccurr$ will become not {\tlower} than $\Cmax$
  (\lstlineref{dbla-replica}{upon-update-complete-ccurr-set}).
  Hence, $r$ will terminate from waiting through the first condition ($C \sqsubset \Ccurr$).
  A contradiction.

  Otherwise, if $\Cmax \sqsubseteq C$, then, by the definition of $\Cmax$,
  $C$ will never be superseded.
  Since $r$ remains correct forever,
  by Lemma~\ref{lem:history-convergence} (History Convergence),
  every correct replica will eventually have $C$ in its local history.
  Since we assume reliable links between processes (see Section~\ref{sec:system-model}),
  every correct replica in $\ireplicas(C)$ will eventually receive $r$'s $\mUpdateRead$ message
  and will send a reply, which $r$ will receive
  (\lstlineref{dbla-replica}{upon-update-read-send-resp}).
  Hence, the waiting on line~\ref{lst:dbla-replica:state-transfer-wait} will eventually terminate through the second condition ($r$ will receive responses from some $Q \in \iquorums(C)$ with the correct sequence number).
  A contradiction.
\end{proof}

Intuitively, the following lemma states that $\Cmax$ is, in some sense, the ``final'' configuration.
After some point every correct process will operate exclusively on $\Cmax$.
No correct process will know about any configuration {\thigher} than $\Cmax$
or ``care'' about any configuration {\tlower} than $\Cmax$.
\begin{lemma} \label{lem:cmax-is-final}
  $\Cmax$ will eventually become the {\thighest} configuration in the local history of each correct process.
\end{lemma}
\begin{proof}
  By Lemma~\ref{lem:history-convergence} (History Convergence), the local histories of all correct processes
  will eventually converge to the same history $h$.
  Let $D = \fMaxElement(h)$.
  Since $\Cmax$ is installed and never superseded, it cannot be {\thigher} than $D$
  \needsrev{(at least one correct replica will always have $\Cmax$ in its local history)}.

  Suppose, for contradiction, that $\Cmax \sqsubset D$.
  In this case, $D$ is never superseded,
  which means that there is a quorum $Q_D \in \iquorums(D)$ that consists
  entirely of forever-correct processes.
  By Lemma~\ref{lem:history-convergence} (History Convergence),
  all replicas in $Q_D$ will eventually \needsrev{have $D$ in their local histories}
  and will try to perform state transfer to it.
  By Lemma~\ref{lem:state-transfer-progress}, they will eventually succeed and install
  $D$---a contradiction with the definition of $\Cmax$.
\end{proof}


\begin{theorem}[\pBLALiveness] \label{the:operational-liveness}
  Our implementation of {\DBLA} satisfies the {\pBLALiveness} property:
  if the total number of verifiable input values is finite,
  every call to $\fPropose(v, \sigma)$ by a forever-correct process eventually returns.
\end{theorem}
\begin{proof}
  Let $p$ be a forever-correct client that invoked $\fPropose(v, \sigma)$.
  By Lemma~\ref{lem:cmax-is-final}, $\Cmax$ will eventually become the
  {\thighest} configuration in the local history of $p$.
  If the client's request will not terminate by the time it learns about $\Cmax$,
  the client will call $\fRefine(\emptyset)$ after it (\lstlineref{dbla-client}{new-history-refine}).
  By Lemma~\ref{lem:cmax-installation},
  $\Cmax$ will eventually be installed by all correct replicas.
  Since it will never be superseded, there will be a quorum of forever-correct replicas.
  Thus, every round of messages from the client will eventually be responded to by a quorum of correct replicas.

  Since the total number of verifiable input values is finite, the client will call $\fRefine$
  only a finite number of times (\lstlineref{dbla-client}{upon-propose-resp-refine}).
  After the last call to $\fRefine$, the client will inevitably receive acknowledgments from a quorum of replicas,
  and will proceed to sending $\mConfirm$ messages
  (\lstlineref{dbla-client}{upon-acks-collected-send-confirm}).
  Again, since there is an available quorum of correct replicas that installed $\Cmax$,
  the client will eventually receive enough acknowledgments and will complete the operation
  (\lstlineref{dbla-client}{propose-after-wait}).
\end{proof}

\begin{theorem}[{\DBLA} liveness] \label{the:dbla-liveness}
  Our implementation satisfies the liveness properties of {\DBLA}:
  {\pBLALiveness},
  {\pDynamicLiveness},
  and {\pInstallationLiveness}.
\end{theorem}
\begin{proof}
  {\pBLALiveness} follows from Theorem~\ref{the:operational-liveness}.
  {\pDynamicLiveness} and {\pInstallationLiveness} follow directly from
  Lemmas~\ref{lem:cmax-is-final} and~\ref{lem:cmax-installation} respectively.
\end{proof}


\section{Max Register}\label{sec:max-register}

\confirmA{%
Our methodology of constructing {\dynamic} and {\reconfigurable} objects
is not limited to lattice agreement.
In this section, we show how to create an atomic Byzantine fault-tolerant {\MaxRegister} in {\dynamic} setting.}

is a distributed object that has two operations: $\fRead()$ and $\fWrite(v, \sigma)$ and
must be parametrized by a boolean function $\fVerifyInputValue(v, \sigma)$.
As before,
we say that $\sigma$ is a \emph{valid certificate for input value}~$v$
iff $\fVerifyInputValue(v, \sigma) = \iTrue$
and that value~$v$ is a \emph{verifiable input value} iff some process knows
$\sigma$ such that $\fVerifyInputValue(v, \sigma) = \iTrue$.
We assume that correct clients invoke $\fWrite(v, \sigma)$ only if $\fVerifyInputValue(v, \sigma) = \iTrue$.
We do not make any assumptions on the number of verifiable input values for this abstraction
(i.e., it can be infinite).

The {\MaxRegister} object satisfies the following three properties:
\begin{itemize}
  \item \textit{\pMRValidity}: if $\fRead()$ returns value $v$ to a correct process, then $v$ is verifiable input value;
  \item \textit{\pMRAtomicity}: if some correct process $p$ completed $\fWrite(v, \sigma)$ or received $v$ from $\fRead()$
    strictly before some correct process $q$ invoked $\fRead()$,
    then the value returned to $q$ must be greater than or equal to $v$;
  \item \textit{\pMRLiveness}: every call to $\fRead()$ and $\fWrite(v, \sigma)$ by a forever-correct process eventually returns.
\end{itemize}
For simplicity, unlike {\ByzantineLatticeAgreement},
our {\MaxRegister} does not provide the $\fVerifyOutputValue(v, \sigma)$ function.

\subsection{{\DynamicMaxRegister} implementation} \label{subsec:dmr-impl}


\begin{algorithm*}
  \caption{{\DynamicMaxRegister}: code for client $p$}
  \labelalg{dmr-client}

  \BeginAlgorithmic
    \Parameters
      \State Lattice of configurations $\Conf$ and the initial configuration $\Cinit \in \Conf$
      \State Set of values $\Values$ and the initial value $\Vinit \in \Values$
      \State Boolean functions $\fVerifyHistory(h, \sigma)$ and $\fVerifyInputValue(v, \sigma)$
    \EndParameters

    \Globals
      \State $\ihistory \subseteq \Conf$, initially $\{ \Cinit \}$  \Comment{local history of this process}
      \State $\iseqNumber \in \mathbb{Z}$, initially $0$  \Comment{used to match requests with responses}
    \EndGlobals

    \AuxiliaryFunctionsInline $\fMaxElement(h)$, $\fFSVerify$ (see Section~\ref{sec:system-model})

    \algspace
    \Operation{$\fRead$}{$ $}                                                                                           \labelline{read}
      \Repeat
        \State let $\Tupple{\ireadOk, \Tupple{v, \sigma}} = \fGet()$
        \State let $\isuccess =$ \textbf{if} $\ireadOk$ \textbf{then} $\fSet(v, \sigma)$ \textbf{else} $\iFalse$        \labelline{read-set}
      \Until{$\isuccess$}
      \State \Return $v$                                                                                                \labelline{read-end}
    \EndOperation

   \algspace
    \Operation{$\fWrite$}{$v$, $\sigma$}                                                                                \labelline{write}
      \Repeat{} let $\isuccess = \fSet(v, \sigma)$
      \Until{$\isuccess$}                                                                                               \labelline{write-end}
    \EndOperation

    \algspace
    \Operation{$\fUpdateHistory$}{$h$, $\sigma$}                                                                        \labelline{update-history}
      \State \RBBroadcast{$\mNewHistory$, $h$, $\sigma$}
    \EndOperation

    \algspace
    \Procedure{$\fSet$}{$v$, $\sigma$}                                                                                      \labelline{set}
      \State $\iseqNumber \gets \iseqNumber + 1$                                                                            \labelline{set-begin}
      \State let $C = \fMaxElement(\ihistory)$
      \State \Send{$\mSet$, $v$, $\iseqNumber$, $C$}{$\ireplicas(C)$}
      \State \WaitFor ($\fMaxElement(\ihistory) \neq C$) $\lor$ (replies from $Q \in \iquorums(C)$ with valid signatures)
      \State \Return $\fMaxElement(\ihistory) \neq C$                                                                       \labelline{set-end}
    \EndProcedure

    \algspace
    \Procedure{$\fGet$}{$ $}                                                                                                \labelline{get}
      \State $\iseqNumber \gets \iseqNumber + 1$                                                                            \labelline{get-begin}
      \State let $C = \fMaxElement(\ihistory)$
      \State \Send{$\mGet$, $\iseqNumber$, $C$}{$\ireplicas(C)$}                                                            \labelline{get-send-request}
      \State \WaitFor ($\fMaxElement(\ihistory) \neq C$) $\lor$ (replies from $Q \in \iquorums(C)$)                         \labelline{get-wait-one}
      \If {$\fMaxElement(\ihistory) \neq C$} \Return $\Tupple{\iFalse, \bot}$
      \Else{} \Return $\Tupple{\iTrue, \text{maximal verifiable input value among received}}$ \EndIf                        \labelline{get-end}
    \EndProcedure

    \algspace
    \UponRBDeliver{$\mNewHistory$, $h$, $\sigma$}{any sender}                                                           \labelline{new-history}
      \If {$\fVerifyHistory(h, \sigma) \land \ihistory \subset h$} $\ihistory \gets h$ \EndIf                           \labelline{history-update}
    \EndHandler
  \EndAlgorithmic
\end{algorithm*}


\begin{algorithm*}
  \caption{{\DynamicMaxRegister}: code for replica $r$}
  \labelalg{dmr-replica}

  \BeginAlgorithmic
    
    \ParametersInline same as in Algorithm~\lstref{dmr-client}.

    \algspace
    \Globals
      \State $\ihistory \subseteq \Conf$, initially $\{ \Cinit \}$                        \Comment{local history of this process}
      \State $\vcurr \in \Values$, initially $\Vinit$
      \State $\sigmacurr \in \Sigma$, initially $\sigma_{init}$
      \State $\Ccurr \in \Conf$, initially $\Cinit$                                       \Comment{current configuration}
      \State $\Cinst \in \Conf$, initially $\Cinit$                                       \Comment{installed configuration}
      \State $\iactiveStateTransfer \in \{\iTrue, \iFalse\}$, initially $\iFalse$
    \EndGlobals

    \algspace
    \AuxiliaryFunctionsInline $\fMaxElement(h)$, $\fFSSign(m, t)$, $\fUpdateFSKeys(t)$ (see Section~\ref{sec:system-model})

    \algspace
    \UponReceive{$\mGet$, $\isn$, $C$}{client $c$}                                                                      \labelline{upon-get}
      \State \WaitFor $C = \Cinst \lor \fMaxElement(\ihistory) \not\sqsubseteq C$                                       \labelline{upon-get-wait}
      \If {$C = \fMaxElement(\ihistory)$}
        \State \Send{$\mGetResp$, $\vcurr$, $\sigmacurr$, $\isn$}{$c$}                       \labelline{upon-get-send-reply}
      \Else{} ignore the message
      \EndIf
    \EndHandler

    \algspace
    \UponReceive{$\mSet$, $v$, $\sigma$, $\isn$, $C$}{client $c$}                                                       \labelline{upon-set}
      \State \WaitFor $C = \Cinst \lor \fMaxElement(\ihistory) \not\sqsubseteq C$                                       \labelline{upon-set-wait}
      \If {$C = \fMaxElement(\ihistory) \land \fVerifyInputValue(v, \sigma)$}
        \If {$v > \vcurr$} $\Tupple{\vcurr, \sigmacurr} \gets \Tupple{v, \sigma}$ \EndIf
        \State \Send{$\mSetResp$, $\fFSSign(\Tupple{c, \isn}, \iheight(C))$, $\isn$}{$c$}
      \Else{} ignore the message
      \EndIf                                                                                                            \labelline{upon-set-end}
    \EndHandler

    \algspace
    \LineCommentx{State transfer}
    \Upon{$\Ccurr \neq \fMaxElement(\{ C \in \ihistory \mid r \in \ireplicas(C) \}) \land \text{not } \iactiveStateTransfer$}                                    \labelline{state-transfer}
      \State Same as for {\DBLA} (\lstlinerangeref{dbla-replica}{state-transfer-begin}{state-transfer-end})
    \EndHandler

    \algspace
    \UponReceive{$\mUpdateRead$, $C$, $\isn$}{replica $r'$}                                                             \labelline{upon-update-read}
      \State \WaitFor $C \sqsubset \fMaxElement(\ihistory)$                                                             \labelline{upon-update-read-wait}
      \State \Send{$\mUpdateReadResp$, $\vcurr$, $\sigmacurr$, $\isn$}{$r'$}
    \EndHandler

    \algspace
    \UponReceive{$\mUpdateReadResp$, $v$, $\sigma$, $\isn$}{replica $r'$}
      \If {$\fVerifyInputValue(v, \sigma) \land v > \vcurr$}
        \State $\Tupple{\vcurr, \sigmacurr} \gets \Tupple{v, \sigma}$ \EndIf  \labelline{upon-read-resp-end}
    \EndHandler

    \algspace
    \UponRBDeliver{$\mNewHistory$, $h$, $\sigma$}{any sender}
      \State Same as for {\DBLA} (\lstlinerangeref{dbla-replica}{new-history}{new-history-end})
    \EndHandler

    \algspace
    \UponURBDeliverIn{$\mUpdateComplete$}{$C$}{quorum $Q \in \iquorums(C)$}                                             \labelline{upon-update-complete}
      \State Same as for {\DBLA} (\lstlinerangeref{dbla-replica}{upon-update-complete}{upon-update-complete-end})
    \EndHandler
  \EndAlgorithmic
\end{algorithm*}

In this section we present our implementation of the \emph{\dynamic} version of the {\MaxRegister} abstraction ({\DynamicMaxRegister} or {\DMR} for short).
Overall, the ``application'' part of the implementation is very similar to the classical ABD algorithm~\cite{ABD},
and the ``{\dynamic}'' part of the implementation is almost the same as in {\DBLA}.

\myparagraph{Client implementation.}
From the client's perspective, the two main procedures are $\fGet()$ and $\fSet(v, \sigma)$
\needsrev{(not to be confused with the $\fRead$ and $\fWrite$ operations)}.
$\fSet(v, \sigma)$ (\lstlinerangeref{dmr-client}{set}{set-end})
is used to store the value on a quorum of replicas of the most recent configuration.
It returns $\iTrue$ iff it manages to receive signed acknowledgments from a quorum of some configuration.
Forward-secure signatures are used to prevent the ``I still work here'' attack.
Since $\fSet$ does not try to read any information from the replicas, it is not susceptible to
the ``slow reader'' attack. 
$\fGet()$ (lines~\ref{lst:dmr-client:get}--\ref{lst:dmr-client:get-end})
is very similar to $\fSet(\ldots)$ and is used to request information from a quorum of replicas of the most recent configuration.
Since we do not provide the $\fVerifyOutputValue(\ldots)$ function,
the replies from replicas are \emph{not signed}
(\lstlineref{dmr-replica}{upon-get-send-reply}).
Therefore, $\fGet()$ is susceptible \needsrev{to both the ``I still work here'' and ``slow reader'' attack} when used alone.
Later in this section we discuss how the invocation of $\fSet(\ldots)$ right after $\fGet()$
(\lstlineref{dmr-client}{read-set})
allows us to avoid these issues.

Operation $\fWrite(v, \sigma)$
(lines~\ref{lst:dmr-client:write}--\ref{lst:dmr-client:write-end})
is used by correct clients to store values in the register.
It simply performs repeated calls to $\fSet(v, \sigma)$ until some call succeeds to reach
a quorum of replicas.
Retries are safe because, as in lattice agreement, write requests to a max-register are idempotent.
Since we assume the total number of verifiable histories to be finite,
only a finite number of retries is possible.

Operation $\fRead()$
(lines~\ref{lst:dmr-client:read}--\ref{lst:dmr-client:read-end})
is used to request the current value from the register,
and it consists of repeated calls to both $\fGet()$ and $\fSet(\ldots)$.
The call to $\fGet()$ is simply used to query information from the replicas.
The call to $\fSet(\ldots)$ is usually called ``the write-back phase'' and serves two purposes here:
\begin{itemize}
  \item It is used instead of the ``confirming'' phase to prevent the ``I still work here'' and the ``slow-reader'' attacks.
    Indeed, if the configuration was superseded during the execution of $\fGet()$,
    $\fSet(\ldots)$ will not succeed because it will not be able to gather a quorum of signed replies
    in the same configuration;

  \item Also, it is used to order the calls to $\fRead()$ and to guarantee the {\pMRAtomicity} property.
    Intuitively, if some correct process successfully completed $\fSet(v, \sigma)$ strictly before some other correct process
    invoked $\fGet()$, the later process will receive a value that is not smaller than $v$
    \needsrev{(unless the ``slow reader'' attack happens)}.
\end{itemize}

\myparagraph{Replica implementation.}

\confirmA{%
The replica implementation (Algorithm~\lstref{dmr-replica})
essentially follows the {\DBLA} guidelines (Algorithms~\lstref{dbla-replica-part-1} and~\lstref{dbla-replica-part-2}),
except that the replica handles client requests specific to {\MaxRegister} (\lstlinerangeref{dmr-replica}{upon-get}{upon-set-end}).
The only other difference is that in handling the $\mUpdateRead$ and $\mUpdateReadResp$ messages (\lstlinerangeref{dmr-replica}{upon-update-read}{upon-read-resp-end}),
the replicas exchange $\vcurr$ and $\sigmacurr$ instead of $\icurrentValues$, as in {\DBLA}.}

\subsection{Proof of correctness}

Since our {\DynamicMaxRegister} implementation uses the same state transfer protocol as {\DBLA}, most proofs from Section~\ref{sec:dbla-correctness-proof} that apply to {\DBLA}, also apply to {\DMR} (with some minor adaptations).
Here we provide only the statements of such theorems, without repeating the proofs.
Then we introduce several theorems specific to {\DMR} and sketch the proofs.
%

\myparagraph{Safety}

\begin{lemma}[Candidate configurations] \label{lem:dmr-candidate-configurations} \leavevmode
  \begin{enumerate}
    \item \label{itm:dmr-cand-in-hist}  Each candidate configuration is present in some verifiable history.
    \item \label{itm:dmr-cand-fin}      There is a finite number of candidate configurations.
    \item \label{itm:dmr-cand-comp}     All candidate configurations are comparable with ``$\sqsubseteq$''. 
  \end{enumerate}
\end{lemma}

\begin{lemma}[{\Tentative} configurations] \label{lem:dmr-tentative-configurations} \leavevmode
  \begin{enumerate}
    \item No correct client will ever make a request to a {\tentative} configuration.
    \item {\Tentative} configurations cannot be installed.
    \item A correct process will never invoke $\fFSVerify$ with timestamp $\iheight(C)$ for any {\tentative} configuration $C$.
    \item A correct replica will never broadcast any message via the uniform reliable broadcast primitive in a {\tentative} configuration.
  \end{enumerate}
\end{lemma}

\begin{lemma} \label{lem:dmr-pivotal-configurations}
  If $C \sqsubseteq \fMaxElement(h)$, where $C$ is a {\pivotal} configuration
  and $h$ is the local history of a correct process,
  then $C \in h$.
\end{lemma}

\begin{theorem}[\pDynamicValidity] \label{the:dmr-dynamic-validity}
  Our implementation of {\DMR} satisfies {\pDynamicValidity}.
  I.e., only a candidate configuration can be installed.
\end{theorem}

\begin{lemma}[Key update] \label{lem:dmr-key-update}
  If a {\pivotal} configuration $C$ is superseded,
  then there is no quorum of replicas in that configuration capable of signing messages with timestamp $\iheight(C)$,
  i.e., $\nexists Q \in \iquorums(C) \textST \forall r \in Q: \ist_r \le \iheight(C)$.
\end{lemma}

We say that a correct client \emph{completes its operation in configuration $C$} iff
at the moment when the client completes its operation,
the {\thighest} configuration in its local history is $C$.

\begin{lemma}[State transfer correctness] \label{lem:dmr-state-transfer-correctness} \leavevmode \\
  If some correct process completed $\fWrite(v, \sigma)$ in $C$
  or received $v$ from $\fRead()$ operation completed in $C$,
  then for each {\tactive} installed configuration $D$ such that $C \sqsubset D$,
  there is a quorum $Q_D \in \iquorums(D)$ such that for each correct replica in $Q_D$:
  $\vcurr \ge v$.
\end{lemma}

The following lemma is the first lemma specific to {\DMR}.
\begin{lemma}[{\pMRAtomicity} in one configuration] \label{lem:dmr-atomicity-in-one-config} \leavevmode \\
  If some correct process $p$ completed $\fWrite(v, \sigma)$ in $C$
  or received $v$ from $\fRead()$ operation completed in $C$
  strictly before some correct process $q$ invoked $\fRead()$
  and $q$ completed its operation in $C$,
  then the value returned to $q$ is greater than or equal to $v$.
\end{lemma}
\begin{proof}
  Recall that $\fRead()$ operation consists of repeated calls to two procedures: $\fGet()$ and $\fSet(\ldots)$.
  If process $q$ successfully completed $\fSet(\ldots)$ in configuration $C$,
  then, by the use of forward-secure signatures,
  configuration $C$ was {\tactive} during the execution of $\fGet()$ that preceded the call to $\fSet$.
  This also means that configuration $C$ was {\tactive} during the execution of $\fSet(v, \sigma)$ by process $p$,
  since it was before process $q$ started executing its request.
  By the quorum intersection property, process $q$ must have received $v$ or a greater value from at least one
  correct replica.
\end{proof}

\begin{theorem}[{\pMRAtomicity}] \label{the:dmr-atomicity}
  Our implementation of {\DMR} satisfies the {\pMRAtomicity} property.
  If some correct process $p$ completed $\fWrite(v, \sigma)$ or received $v$ from $\fRead()$
  strictly before some correct process $q$ invoked $\fRead()$,
  then the value returned to $q$ must be greater than or equal to $v$
\end{theorem}
\begin{proof}
  Let $C$ (resp., $D$) be the {\thighest} configuration in $p$'s (resp., $q$'s) local history when it completed its request.
  Also, let $v$ (resp., $u$) be the value that $p$ (resp., $q$) passed to the last call to $\fSet(\ldots)$
  (note that both $\fRead()$ and $\fWrite(\ldots)$ call $\fSet(\ldots)$).

  If $C = D$, then $u \ge v$ by Lemma~\ref{lem:dmr-atomicity-in-one-config}.

  \needsrev{%
  Suppose, for contradiction, that $D \sqsubset C$.
  Since correct replicas do not reply to user requests in a configuration until this configuration is installed
  (\lstlineref{dmr-replica}{upon-get-wait}), configuration $C$ had to be installed before $p$ completed its request.
  By Lemma~\ref{lem:dmr-key-update} (Key Update), 
  this would mean that $q$ would not be able to complete $\fSet(\ldots)$ in $D$---a contradiction.}

  \needsrev{The remaining case is when $C \sqsubset D$.}
  In this case, by Lemma~\ref{lem:dmr-state-transfer-correctness},
  the quorum intersection property,
  and the use of forward-secure signatures in $\fSet(\ldots)$,
  $q$ received $v$ or a greater value from at least one
  correct replica during the execution of $\fGet()$.
  Therefore, in this case $u$ is also greater than or equal to $v$.
\end{proof}

\begin{theorem}[{\DMR} safety] \label{the:dmr-safety}
  Our implementation satisfies the safety properties of {\DMR}:
  {\pMRValidity},
  {\pMRAtomicity},
  and {\pDynamicValidity}.
\end{theorem}
\begin{proof}
  {\pMRValidity} follows directly from the implementation:
  correct clients only return verifiable input values from $\fGet()$ (\lstlineref{dmr-client}{get-end}).
  {\pMRAtomicity} follows directly from Theorem~\ref{the:dmr-atomicity}.
  {\pDynamicValidity} follows from Theorem~\ref{the:dmr-dynamic-validity}.
\end{proof}

\myparagraph{Liveness}

\begin{lemma}[History Convergence] \label{lem:dmr-history-convergence}
  Local histories of all correct processes will eventually become identical.
\end{lemma}

Recall that the \emph{maximal installed configuration} is the {\thighest} installed configuration
and is denoted by $\Cmax$ \needsrev{(see Definition~\ref{def:cmax} and Lemma~\ref{lem:cmax-existence} in Section~\ref{sec:dbla-correctness-proof})}.

\begin{lemma}[$\Cmax$ installation] \label{lem:dmr-cmax-installation}
  The maximal installed configuration will eventually be installed by all correct replicas.
\end{lemma}

\begin{lemma}[State transfer progress] \label{lem:dmr-state-transfer-progress}
  State transfer executed by a forever-correct replica always terminates.
\end{lemma}

\begin{lemma} \label{lem:dmr-cmax-is-final}
  $\Cmax$ will eventually become the {\thighest} configuration in the local history of each correct process.
\end{lemma}


\begin{theorem}[\pMRLiveness] \label{the:dmr-mr-liveness}
  Our implementation of {\DMR} satisfies the {\pMRLiveness} property:
  every call to $\fRead()$ and $\fWrite(v, \sigma)$ by a forever-correct process eventually returns.
\end{theorem}
\begin{proof}
  Let $p$ be a forever-correct client that invoked $\fRead()$ or $\fWrite(\ldots)$.
  By Lemma~\ref{lem:cmax-is-final}, $\Cmax$ will eventually become the
  {\thighest} configuration in the local history of $p$.
  If the client's request does not terminate by the time the client learns about $\Cmax$,
  the client will restart the request in $\Cmax$.
  Since $\Cmax$ will eventually be installed by all correct replicas and will never
  be superseded, there will be a quorum of forever-correct replicas,
  and $p$ will be able to complete its request there.
\end{proof}

\begin{theorem}[{\DMR} liveness] \label{the:dmr-liveness}
  Our implementation satisfies the liveness properties of {\DMR}:
  {\pMRLiveness},
  {\pDynamicLiveness},
  and {\pInstallationLiveness}.
\end{theorem}
\begin{proof}
  {\pMRLiveness} follows from Theorem~\ref{the:dmr-mr-liveness}.
  {\pDynamicLiveness} and {\pInstallationLiveness} follow directly from
  Lemmas~\ref{lem:dmr-cmax-is-final} and~\ref{lem:dmr-cmax-installation} respectively.
\end{proof}


\end{document}